\documentclass[prd,preprint,superscriptaddress,amsmath,amssymb,nofootinbib]{revtex4}
\usepackage{graphicx,color,slashed}

 \usepackage{epstopdf}
 \usepackage{slashed}
 \usepackage{ulem}

 \usepackage[bookmarks=true,bookmarksnumbered=true,bookmarkstype=toc]{hyperref} 
\newcommand{\lsim}{\raise0.3ex\hbox{$\;<$\kern-0.75em\raise-1.1ex\hbox{$\sim\;$}}}
\newcommand{\gsim}{\raise0.3ex\hbox{$\;>$\kern-0.75em\raise-1.1ex\hbox{$\sim\;$}}}

\begin{document}

{\begin{flushright}
{KIAS-P18019} \\
{UME-PP-008}
\end{flushright}}

\title{ Searching for scalar boson decaying into light $Z'$ boson \\
at collider experiments in $U(1)_{L_\mu - L_\tau}$ model }

\author{Takaaki Nomura}
\email{nomura@kias.re.kr}
\affiliation{School of Physics, KIAS, Seoul 130-722, Korea}

\author{Takashi Shimomura}
\email{shimomura@cc.miyazaki-u.ac.jp}
\affiliation{Faculty of Education, Miyazaki University, Miyazaki, 889-2192, Japan}

\date{\today}

\begin{abstract}
We study a model with $U(1)_{L_\mu - L_\tau}$ gauge symmetry and discuss collider searches for 
a scalar boson, which breaks $U(1)_{L_\mu - L_\tau}$ symmetry spontaneously, decaying into 
light $Z'$ gauge boson.
In this model, the new gauge boson, $Z'$, with a mass lighter than $\mathcal{O}(100)$ MeV,  
plays a role in explaining the anomalous magnetic moment of muon via one-loop contribution. 
For the gauge boson to have such a low mass, the scalar boson, $\phi$ with $\mathcal{O}(100)$ GeV mass appears associated with the symmetry breaking. 
We investigate experimental constraints on $U(1)_{L_\mu - L_\tau}$ gauge coupling, kinetic mixing, and mixing between the SM Higgs and $\phi$.
Then collider search is discussed considering $\phi$ production followed by decay process $\phi \to Z' Z'$ at the large hadron collider and the international linear collider. 
We also estimate discovery significance at the linear collider taking into account relevant kinematical cut effects.
  
\end{abstract}
\maketitle

\section{Introduction}
The standard model (SM) of particle physics has been describing phenomena over the wide range of energy scale from 
eV to TeV scale. Despite of such enormous success, the anomalous magnetic moment of the muon, $(g-2)_\mu$, shows 
a long-standing discrepancy between experimental observations \cite{Bennett:2006fi, Patrignani:2016xqp}
and theoretical predictions \cite{Davier:2010nc, Jegerlehner:2011ti, Hagiwara:2011af, Aoyama:2012wk}, 
\begin{align}
\Delta a_\mu \equiv \Delta a_\mu^{\mathrm{exp}} - \Delta a_\mu^{\mathrm{th}} = (28.8 \pm 8.0) \times 10^{-10},
\end{align}
where $a_\mu = (g-2)_\mu/2$. This difference reaches to $3.6\sigma$ deviation from the prediction and 
seems not to be resolved within the SM. The on-going and forthcoming experiments will verify the discrepancy with high 
statistics, which will reduce the uncertainties by a factor of four \cite{Grange:2015fou, Saito:2012zz}. 
Then, when the discrepancy is confirmed by the these experiments, it must be a firm evidence of physics beyond the SM. 

Many extensions of the SM have been proposed to resolve the discrepancy so far (See for a review \cite{Lindner:2016bgg}). Among them,  one of the 
minimal extensions is to add a new $U(1)$ gauge symmetry to the SM. When muon is charged under the symmetry, 
the deviation of $(g-2)_\mu$ can be explained by a new contribution from the associated gauge boson of the symmetry 
through loop diagrams. 
The $L_\mu - L_\tau$ gauge symmetry is particularly interesting in this regard because it is anomaly free extension and can also 
explain the neutrino mass and mixings simultaneously~\cite{He:1990pn, Foot:1994vd, Asai:2017ryy}. In this model, it was shown in refs.~\cite{Gninenko:2001hx, Baek:2001kca,Ma:2001md} that the deviation 
of $(g-2)_\mu$ can be resolved when the gauge boson mass is of order $10$ MeV and the gauge coupling constant 
is of order $10^{-4}$. Such a light and weakly interacting gauge boson is still allowed from experimental searches performed in past. 
Interestingly, it was also shown that the gauge boson with similar mass and gauge coupling can also explain 
the deficit of cosmic neutrino flux reported by IceCube collaboration~\cite{Aartsen:2014gkd, Araki:2014ona, Kamada:2015era, DiFranzo:2015qea, Araki:2015mya}. 
Many experimental searches have been prepared and on-going for such a light particles in meson decay 
experiment \cite{Banerjee:2017hhz}, beam dump experiment \cite{Anelli:2015pba} and  electron-positron collider 
experiment \cite{Abe:2010gxa}. Theoretical studies on search strategy at collider experiment are also proposed (see e.g. \cite{Heeck:2011wj, Harigaya:2013twa, delAguila:2014soa,delAguila:2015vza} 
for $L_\mu - L_\tau$ model~\footnote{In these analyses, $Z'$ mass is considered to be $\mathcal{O}(10)$-$\mathcal{O}(100)$ GeV and $Z'$ can decay into charged leptons $\mu^+\mu^-(\tau^+\tau^-)$ providing four charged lepton signals.   }).

As mentioned above, the $L_\mu - L_\tau$ gauge boson has a mass, hence the symmetry must be broken. This implies that 
at least one new complex scalar, which is singlet under the SM gauge group, should exist to break the symmetry and 
give a mass to the gauge boson. Then, from the gauge symmetry, there must exist an interaction of two gauge bosons and one 
real scalar by replacing the scalar field with its vacuum expectation value (VEV).  Since this interaction is generated after 
the symmetry breaking, the confirmation of the interaction by experiments is a crucial to identify the model.
The VEV of the scalar can be estimated as about $10$-$100$ GeV 
from the gauge boson mass and the gauge coupling. Thus, naively one can expect that the physical CP-even scalar emerging after the 
symmetry breaking has a mass of the same order. Such a heavy scalar can not be directly searched at low energy experiments, 
and hence should be searched at high energy collider experiments, i.e. the Large Hadron Collider (LHC) experiment and 
future International Linear Collider (ILC) experiment \cite{Baer:2013cma, Fujii:2017vwa}.
In this paper, we study signatures for the scalar as well as the light gauge boson using
 $Z'$-$Z'$-$\phi$ vertex at the LHC experiment and $Z$-$Z$-$\phi$ vertex at the ILC experiments. 

This paper is organized as follows. In section II, we briefly review the minimal gauged $L_\mu - L_\tau$ model and give the 
partial decay widths of the scalar  and gauge bosons. In section III, we show the allowed parameter space of the model.
Then we show our results on the signatures of the scalar and the gauge boson production at the LHC and ILC experiments 
in section IV. Section V is devoted to the summary and discussion.

\begin{center} 
\begin{table}
\begin{tabular}{|c||c|c||c|c|c|c|c|c|}\hline\hline  
& \multicolumn{2}{|c||}{Scalar} & \multicolumn{6}{c|}{Lepton} \\ \hline \hline
& ~$H$~ & ~$\varphi$ ~ &  ~$L_e$~ & ~$L_\mu$~ & ~$L_\tau$~ & ~$e_R$~ & ~$\mu_R$~ & ~$\tau_R$~ \\\hline 
$SU(2)_L$ & $\bf{2}$  & $\bf{1}$  &\bf{2} & $\bf{2}$ & $\bf{2}$ & $\bf{1}$ & $\bf{1}$ & \bf{1} \\\hline 
$U(1)_Y$ & $\frac{1}{2}$  & $0$ & $-\frac{1}{2}$ & $-\frac{1}{2}$ & $-\frac{1}{2}$ & $-1$ & $-1$ & $-1$ \\\hline
$U(1)_{L_\mu - L_\tau}$ & $0$ & $1$ & $0$ & $1$ & $-1$ & $0$ & $1$ & $-1$   \\\hline
\end{tabular}
\caption{Contents of scalar fields 
and their charge assignments under $SU(2)_L\times U(1)_Y \times U(1)_{L_\mu -L_\tau}$.}
\label{tab:1}
\end{table}
\end{center}
\section{Model}

We begin our discussions with reviewing a model with gauged $U(1)_{L_\mu - L_\tau}$ symmetry under which 
muon ($\mu$) and tau ($\tau$) flavor leptons are charged among the SM leptons. 
As a minimal setup, we introduce a SM singlet scalar field $\varphi$ to break the $L_\mu - L_\tau$ symmetry spontaneously. 
The gauge charge assignment for the lepton and scalar fields are given in Table~\ref{tab:1}, and the quark sector is 
the same as that of the SM. In the table, $L_{e},~L_{\mu},~L_{\tau}$ and $e_R,~\mu_R,~\tau_R$ denote the left and right-handed leptons, 
and $H$ denotes the $SU(2)_L$ doublet scalar field, respectively. 
The Lagrangian of the model is given by
\begin{align}
\mathcal{L} =& \mathcal{L}_{\rm SM} +|D_\mu \varphi|^2 - V - \frac{1}{4} Z'_{\mu \nu} Z'^{\mu \nu} - \frac{\epsilon}{2} B_{\mu \nu} Z'^{\mu \nu} + g' Z'_\mu J^\mu_{Z'}, \label{eq:Lagrangian} \\
J^\mu_{Z'} =& \bar L_\mu \gamma^\mu L_\mu + \bar \mu_R \gamma^\mu \mu_R - \bar L_\tau \gamma^\mu L_\tau - \bar \tau_R \gamma^\mu \tau_R, \\
V = &  -\mu_H^2 H^\dagger H - \mu_\varphi^2 \varphi^* \varphi + \frac{\lambda_H}{2} (H^\dagger H)^2 + \frac{\lambda_\varphi}{2} (\varphi^* \varphi)^2 + \lambda_{H \varphi} (H^\dagger H)(\varphi^* \varphi), \label{eq:scalar-potential}
\end{align}  
where $\mathcal{L}_{\rm SM}$, $J_{Z'}$ and $V$ represent the SM Lagrangian, the $U(1)_{L_\mu- L_\tau}$ current and 
the scalar potential, respectively. 
The gauge fields and its field strengths corresponding to $U(1)_{L_\mu - L_\tau}$ and $U(1)_Y$ are denoted 
by $Z'$ and $B$. In Eq.\eqref{eq:Lagrangian}, $D_\mu = \partial_\mu - i g' Z'_\mu$ is the covariant derivative, 
and $g'$ and $\epsilon$ represent the $L_{\mu }- L_{\tau}$ gauge coupling constant and 
the kinetic mixing parameter, respectively. In the following discussions, we assume that the quartic couplings 
of the scalar fields, $\lambda_H,~\lambda_\varphi$ and $\lambda_{H\varphi}$, are positive to avoid runaway directions. 
In Eq.\eqref{eq:scalar-potential}, $\mu_H^2$ and $\mu_\phi^2$ are the tachyonic masses of $H$ and $\varphi$.

The scalar fields $H$ and $\varphi$ can be expanded as
\begin{equation}
H = \begin{pmatrix} H^+ \\ 
\frac{1}{\sqrt{2}} (v + \tilde H + i A) \end{pmatrix}, \quad \varphi = \frac{1}{\sqrt{2}} (v_\varphi + \tilde \phi + i a),
\label{eq:scalars}
\end{equation} 
where $H^+$, $A$ and $a$ are massless Nambu-Goldstone bosons which should be absorbed by the gauge bosons $W^+$, $Z$ and $Z'$, 
while $\tilde{H}$ and $\tilde{\phi}$ represent the physical CP-even scalar bosons.

The VEVs of the scalar fields, $v$ and $v_\varphi$, are obtained from the stationary conditions 
$\partial V/ \partial v = \partial V/ \partial v_\varphi = 0$;
\begin{equation}
v = \sqrt{\frac{2 (\lambda_\varphi \mu_H^2 - \lambda_{H \varphi} \mu_\varphi^2 )}{\lambda_H \lambda_\varphi - \lambda_{H \varphi}^2 }}, \quad
v_\varphi = \sqrt{\frac{2 ( \lambda_H \mu_\varphi^2 - \lambda_{H \varphi} \mu_H^2) }{\lambda_H \lambda_\varphi - \lambda_{H \varphi}^2 }}. 
\end{equation}
Without loss of generality, the VEVs are taken to be real-positive by using the degree of freedom of the gauge symmetries 
to rotate the scalar fields.
Inserting Eq.\eqref{eq:scalars} into Eq.\eqref{eq:scalar-potential}, the squared mass terms for CP-even scalar bosons are given by
\begin{equation}
\mathcal{L} \supset \frac{1}{4} \begin{pmatrix} \tilde H \\ \tilde \phi \end{pmatrix}^T \begin{pmatrix} \lambda_H v^2 & \lambda_{H \varphi} v v_\varphi \\  \lambda_{H \varphi} v v_\varphi  & \lambda_\varphi v_\varphi^2 \end{pmatrix} \begin{pmatrix} \tilde H \\ \tilde \phi \end{pmatrix}.
\end{equation} 
The above squared mass matrix can be diagonalized by an orthogonal matrix. The mass eigenvalues are given by
\begin{equation}
m_{h,\phi}^2 = \frac{\lambda_H v^2 +\lambda_\varphi v_\varphi^2 }{4} \pm \frac{1}{4} \sqrt{\left( \lambda_H v^2 -\lambda_\varphi v_\varphi^2 \right)^2 + 4 \lambda_{H \varphi}^2 v^2 v_\varphi^2 }, 
\end{equation}
and the corresponding mass eigenstates $h$ and $\phi$ are obtained as   
\begin{equation}
\begin{pmatrix} h \\ \phi \end{pmatrix} = \begin{pmatrix} \cos \alpha & \sin \alpha \\ - \sin \alpha & \cos \alpha \end{pmatrix} \begin{pmatrix} \tilde H \\ \tilde \phi \end{pmatrix}, \quad
\tan 2 \alpha = \frac{2 \lambda_{H \varphi} v v_\varphi}{\lambda_H v^2 - \lambda_\varphi v_\varphi^2},
\label{eq:scalar-mass-fields}
\end{equation}
where $\alpha$ is the mixing angle. When $\alpha \ll 1$, $h$ is identified as the SM-like Higgs boson.
Note that the scalar quartic couplings, $\lambda_\varphi$ and $\lambda_{H\varphi}$, 
are smaller than unity in our discussion. In fact, the typical order of these couplings are $\mathcal{O}(10^{-2})$ 
and $\mathcal{O}(10^{-3})$, respectively, when we take $\sin\alpha=0.05$ and $m_\phi=\mathcal{O}(100)$ GeV, $m_{Z'} = 100$ MeV. Therefore the perturbative unitarity and stability of the potential are maintained at least 
up to $10$ TeV. 

After the spontaneous breaking of the electroweak and $L_\mu -L_\tau$ symmetries, the gauge bosons acquire masses. 
The neutral components of the gauge bosons mix each other due to the kinetic mixing 
while the charged ones remain the same as those of the SM. 
Assuming $\epsilon \ll 1$, the mass eigenvalues of the neutral components, $Z_{1,2,3}$, are obtained after diagonalizing the mass term 
as well as the kinetic term,
\begin{subequations}
\begin{align}
m_{Z_1}^2 &= 0, \\
m_{Z_2}^2  &= m_Z^2 (1 - 2 \epsilon^2 \sin^2 \theta_W) + \mathcal{O}(\epsilon^4 m_Z^2), \\
m_{Z_3}^2 &=  m_{Z'}^2 + \mathcal{O}(\epsilon^6 m_Z^2),
\end{align}
\label{eq:gauge-masses}
\end{subequations}
where $m_Z$ and $\theta_W$ are the $Z$ boson mass and the Weinberg angle in the SM, respectively, and 
\begin{align}
m_{Z'} = g' v_\varphi
\end{align}
The corresponding mass eigenstates of the gauge bosons are given by
\begin{subequations}
\begin{align}
Z_1^\mu &= A^\mu,\\
Z_2^\mu &\simeq Z^\mu, \\
Z_3^\mu &\simeq Z'^\mu - \epsilon \sin \theta_W Z^\mu,
\end{align}
\label{eq:gauge-mass-fields}
\end{subequations}
up to $\mathcal{O}(\epsilon^2)$. Thus, $Z_1$ is the photon, and $Z_2$ and $Z_3$ are almost $Z$ and $Z'$, respectively. 
We denote $Z_1$ and $Z_2$ as $Z$ and $Z'$ in the rest of this paper. 
Note that $\rho$-parameter in our model is shifted from 1 as 
\begin{equation}
\rho = \frac{m_{Z}^2}{m_{Z_2}^2} \simeq 1 + 2 \epsilon^2 \sin^2 \theta_W \simeq 1 + \frac{1}{2} 10^{-6} 
\left( \frac{\epsilon}{10^{-3}} \right)^2,
\end{equation}
where the experimental observation is given by $\rho = 1.0004^{+0.0003}_{-0.0004}$~\cite{PDG} with $2 \sigma$ error. Thus we can avoid the constraint from $\rho$-parameter for $\epsilon \lesssim 3.7 \times 10^{-2}$.

The Yukawa and gauge interactions of the SM fermions and $\phi$ in mass-basis are given by 
\begin{align}
\label{eq:intV}
\mathcal{L} \supset &  \sum_{f} \frac{m_f}{v} \sin \alpha \phi \bar f f + \frac{m_{Z'}^2}{v_\varphi} \cos \alpha \phi Z'_\mu Z'^\mu 
+ \frac{m_Z^2}{v} \sin \alpha \phi Z_\mu Z^\mu + \frac{2 m_W^2}{v} \sin \alpha \phi W^+_\mu W^{-\mu}  \nonumber \\
&+ Z'_\mu (-e \epsilon \cos \theta_W J_{\rm EM}^\mu + g' J_{Z'}^\mu) + \mathcal{O}(\epsilon^2),
\end{align}
where $m_f$ and $J_{\rm EM}$ represent the mass of the fermions $f$ and the electromagnetic currents of the SM, and $e$ and $\theta_W$ 
are the electric charge of the proton and the Weinberg angle, respectively. 
In Eq.\eqref{eq:intV}, 
the interactions between $Z'$ and $J_{\rm EM}$ are induced through the kinetic mixing~\footnote{Then $Z'$ interaction is flavor diagonal and $K$-meson and $B$-meson physics do not give significant constraints to the $Z'$ coupling and mass.}. 
In the LHC and lepton collider experiments, the scalar $\phi$ can be mainly produced via the gluon fusion and associate $Z$ 
production processes. One can see from Eq.\eqref{eq:intV} that the relevant interactions are proportional to the scalar 
mixing, $\sin\alpha$. Therefore the production cross section increases as $\sin\alpha$ becomes larger.

For the SM-like Higgs boson, the gauge and scalar interactions in mass-basis are also obtained by inserting Eq.\eqref{eq:scalar-mass-fields} 
into the Lagrangian. The relevant interactions in our discussions are given by
\begin{align}
\label{eq:intH}
\mathcal{L} \supset  \frac{m_{Z'}^2}{v} \sin\alpha h Z'_\mu Z'^\mu - \frac{1}{2} g_{h \phi \phi} h \phi \phi
+ \mathcal{O}(\epsilon),
\end{align}
where $g_{h \phi\phi}$ is the constant given by
\begin{align}
g_{h \phi\phi} &= 3 \sin\alpha \cos\alpha ( \lambda_H v \sin\alpha + \lambda_\varphi v_\varphi \cos\alpha ) \nonumber \\
&+\lambda_{H\varphi} (v \cos^3\alpha + v_\varphi \sin^3\alpha  - 2 v_\varphi \sin\alpha \cos^2 \alpha  
   - 2 v \sin^2 \alpha \cos \alpha).
\end{align}
There exist other gauge and scalar-self interactions involving $h$. However, those are negligible 
when the mixing angle $\alpha$ and the kinetic mixing parameter $\epsilon$ is much smaller than the unity.
Note that $\lambda_{H \varphi}$ is written in terms of $\alpha$ from Eq.\eqref{eq:scalar-mass-fields}.
Therefore $g_{h \phi\phi}$ becomes proportional to $\alpha$ when $\alpha$ is small enough.

In the end of this section, we show the decay widths of $\phi$, $Z'$ and $h$. 
As we will explain in the next section, we focus our discussions on the situation that the $Z'$ gauge boson has a mass lighter than 
$2m_\mu$, while the scalar boson $\phi$ has a mass of order $10$-$100$ GeV. 
Thus, the $\phi$ can decay into $Z'$ as well as the SM fermions and the gauge bosons.
 The partial decay widths of $\phi$ are given by
\begin{align}
\Gamma_{\phi \to Z' Z'} & = \frac{{g'}^2 \cos^2 \alpha}{8 \pi}  \frac{m_{Z'}^2}{m_\phi} \sqrt{1 - \frac{4 m_{Z'}^2}{m_\phi^2}} \left( 2 + \frac{m_\phi^4}{4 m_{Z'}^4} \left( 1 - \frac{2 m_{Z'}^2 }{m_\phi^2} \right)^2 \right), \\ 
\Gamma_{\phi \to f \bar f} &= \frac{m_\phi}{8 \pi} \left( \frac{m_f}{v} \right)^2 \sin^2 \alpha \left( 1 - \frac{4 m_f^2}{m_\phi^2} \right)^{\frac32}, \\
\Gamma_{\phi \to Z Z(W^+W^-)} & = \frac{\sin^2 \alpha}{8 \pi} \frac{m_{Z(W)}^3}{v^2} \frac{m_{Z(W)}}{m_\phi} \sqrt{1 - \frac{4 m_{Z(W)}^2}{m_\phi^2}} \left( 2 + \frac{m_\phi^4}{4 m_{Z(W)}^4} \left( 1 - \frac{4 m_{Z(W)}^2 }{m_\phi^2} \right)^2 \right), 
\end{align}
where $m_{Z,W^\pm}$ are the mass of the gauge bosons, respectively. 
Here we have assumed final states are on-shell. 
It is important to mention that $\phi$ dominantly decays 
into $Z' Z'$ when $Z'$ mass is light since its partial decay width is enhanced by $m_\phi^4/m_{Z'}^4$ factor.

\begin{figure}[t]
\begin{center}
\includegraphics[width=70mm]{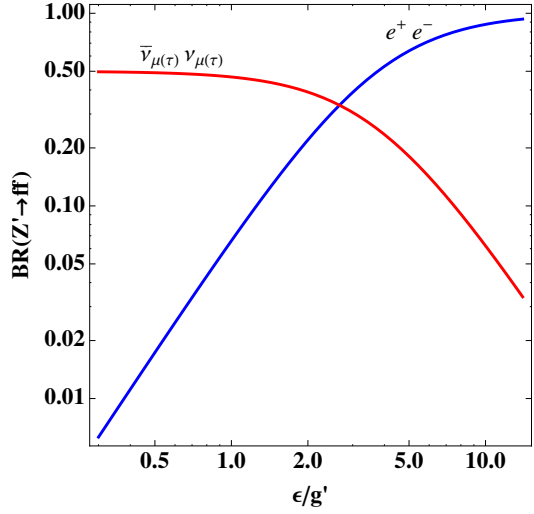} 
\caption{$BR(Z' \to ff)$ as a function of $\epsilon/g'$ where red and blue lines correspond to $\bar \nu_{\mu,\tau} \nu_{\mu,\tau}$ and $e^+e^-$ mode respectively. The mass of $Z'$ is fixed to $100$ MeV.} 
  \label{fig:BR}
\end{center}\end{figure}

The $Z'$ boson can decay into $\bar \nu_{\mu,\tau} \nu_{\mu, \tau}$ or $e^+ e^-$ modes because $m_{Z'} < 2 m_\mu$.
Then the partial decay widths of $Z'$ are obtained as
\begin{align}
\Gamma_{Z' \to \nu \bar \nu} = & \frac{g'^2}{24 \pi} m_{Z'}, \\
\Gamma_{Z' \to e^+ e^-} =& \frac{e^2 \epsilon^2 \cos^2 \theta_W}{12 \pi} m_{Z'} \left( 1 + 2 \frac{m_e^2}{m_{Z'}^2} \right) \sqrt{1 - \frac{4 m_e^2}{m_{Z'}^2}},
\end{align}
where we have ignored the neutrino masses and mixing.
The branching ratio (BR) can be parametrized by the ratio of $L_\mu - L_\tau$ gauge coupling and 
kinetic mixing parameter, $\epsilon/g'$. We show $BR(Z' \to ff)$ as a function of $\epsilon/g'$ in Figure~\ref{fig:BR} where red and blue 
curves respectively correspond to $\bar \nu_{\mu,\tau} \nu_{\mu,\tau}$ and $e^+e^-$ mode. 
The mass of $Z'$ is fixed to $100$ MeV, however the branching ratio is almost independent of the $Z'$ mass 
when $m_{Z'} \gg m_e$. 
It is seen in Fig.~\ref{fig:BR} that $Z'$ mainly decays into neutrinos for $\epsilon/g'<1$.
For later use, the branching ratio is about $0.07$ for $\epsilon/g' = 1$.

The SM-like Higgs boson can decay into not only $Z'$ but also $\phi$ when $m_\phi < m_h/2$. The partial widths of these 
decays are given by
\begin{subequations}
\begin{align}
\Gamma_{h \rightarrow Z' Z'} & = \frac{{g'}^2 \sin^2 \alpha}{8 \pi}  \frac{m_{Z'}^2}{m_{h}} \sqrt{1 - \frac{4 m_{Z'}^2}{m_{h}^2}}
\left( 2 + \frac{m_{h}^4}{4 m_{Z'}^4} \left( 1 - \frac{2 m_{Z'}^2 }{m_{h}^2} \right)^2 \right), 
\label{eq:gamma-HZpZp}\\ 
\Gamma_{h \rightarrow \phi \phi} &=\frac{g_{h \phi\phi}^2}{32\pi m_{h}} \sqrt{1 - \left(\frac{2m_\phi}{m_{h} } \right)^2}.
\label{eq:gamma-Hphiphi}
\end{align}
\label{eq:gamma-Hinvis}
\end{subequations}

\begin{figure}[t]
\begin{center}
\includegraphics[width=70mm]{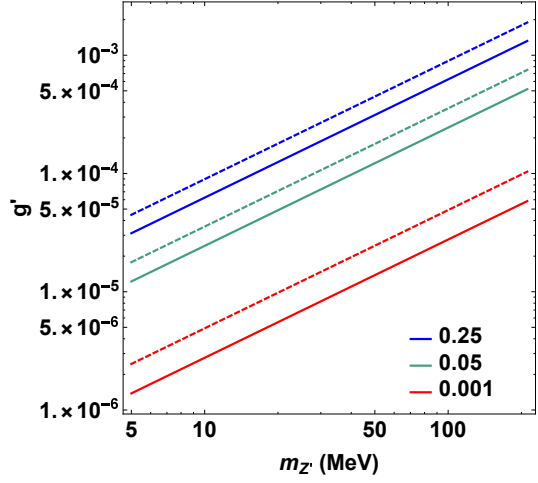} 
\caption{The branching ratio of the Higgs invisible decays, $h \rightarrow Z'Z'$ and $h \rightarrow \phi \phi$.} 
  \label{fig:higgs-invisible}
\end{center}\end{figure}

As we mentioned above, $\phi$ dominantly decays into $Z'$, and $Z'$ mainly decays into neutrinos for $\epsilon/g' < 1$.
Therefore these decays are invisible.
The branching ratio of the invisible decays in our model is given by
\begin{align}
BR(h \rightarrow \mathrm{invisibles}) \equiv 
\frac{\Gamma_{h \rightarrow Z'Z'} + \Gamma_{h \rightarrow \phi \phi}}{\Gamma_{\mathrm{SM}} + \Gamma_{h \rightarrow Z'Z'} + \Gamma_{h \rightarrow \phi \phi}},
\label{eq:higgs-invisible-br}
\end{align}
where $\Gamma_{\mathrm{SM}}$ is the total width of the Higgs boson in the SM\footnote{Another invisible decay of the Higgs boson 
$h \rightarrow ZZ^\ast \rightarrow \nu\nu \bar{\nu} \bar{\nu}$, exists within the SM. 
The partial width of this decay is about $4.32$ keV \cite{deFlorian:2016spz}, and it is much smaller than the widths 
of $h \rightarrow Z'Z'/\phi\phi$ in our parameter region. Thus, we have neglected this.}.
When $m_\phi \geq m_{h}/2$,  $\Gamma_{h \rightarrow \phi\phi}$ should be dropped in Eq.\eqref{eq:higgs-invisible-br}. 
The invisible decay of the Higgs boson has been searched at the LHC experiment in the production 
via gluon fusion \cite{Khachatryan:2016whc}, vector boson fusion 
\cite{Khachatryan:2016whc, Chatrchyan:2014tja, Aad:2015txa, Aad:2015pla},
and in association with a vector boson 
\cite{Khachatryan:2016whc, Chatrchyan:2014tja, Aad:2015pla, Aad:2014iia, Aad:2015uga, Aaboud:2017bja}.
We employ $BR(h \rightarrow \mathrm{invisibles}) \leq 0.25$ given in \cite{Aad:2015pla}.
In Figure~\ref{fig:higgs-invisible}, the branching ratio is shown in $m_{Z'}$-$g'$ plane.
The blue, green and red lines correspond to $BR(h \rightarrow \mathrm{invisibles}) =0.25, 0.05$ and $0.001$, respectively. 
The scalar mass is taken as $m_\phi = 30$ GeV (solid) and $m_\phi \geq m_h/2$ (dashed), and the scalar mixing 
is fixed to $\sin\alpha = 0.03$ for reference. The SM-like Higgs mass and its total decay width is taken as $125$ GeV \cite{Aad:2015zhl} 
and $4.07$ MeV \cite{Dittmaier:2011ti}, respectively.
From the figure, we can see that $g'$ should be smaller than $2 \times 10^{-3}$ for $m_{Z'} \leq 2 m_\mu$, 
to avoid the upper bound from the LHC experiment. 
This region of $g'$ is consistent with the favored region to resolve $(g-2)_\mu$ discrepancy.

\section{Allowed parameter space}
In this section, we show the allowed parameter space of $g'$, $\epsilon$ and $m_{Z'}$, $\alpha$. The parameters of $Z'$ are 
tightly constrained by experiments such as beam dump experiments \cite{Riordan:1987aw, Blumlein:2011mv}, 
meson decay experiments  \cite{Adler:2004hp, Artamonov:2008qb, Batley:2015lha, Banerjee:2016tad,Banerjee:2017hhz}, 
neutrino-electron scattering measurements \cite{Bellini:2011rx}, electron-positron collider experiment \cite{Lees:2014xha, TheBABAR:2016rlg}, neutrino trident production process \cite{Geiregat:1990gz,Mishra:1991bv}. 
A hadron collider experiment such as the LHC also constrains the gauge interaction for heavier $Z'$ region~\cite{Harigaya:2013twa, delAguila:2014soa, delAguila:2015vza} although we will not discuss such a heavy $Z'$. 
The parameters can be further constrained 
by requiring that the $Z'$ gauge boson gives enough contributions to $(g-2)_\mu$.

As we mentioned in the introduction, the deviation of $(g-2)_\mu$ between the experimental observations and 
the theoretical prediction are 
\begin{align}
12.8~(4.8) \leq \Delta a_\mu^{Z'} \times 10^{10} \leq 44.8~(52.8).
\end{align}
within $2\sigma~(3\sigma)$.  The contribution from $Z'$ to the anomalous magnetic moment is given by
\begin{align}
\Delta a_\mu^{Z'} = \frac{(g'+\epsilon e \cos\theta_W)^2}{8 \pi^2} \int^1_0 dx \frac{2 m_\mu^2 x^2 (1-x)}{x^2 m_\mu^2 + (1-x)m_{Z'}^2}.
\label{eq:delta-mu-g-2}
\end{align}
The favored region of the gauge coupling and the $Z'$ mass 
to explain the deviation were studied in \cite{Altmannshofer:2014pba, Araki:2017wyg, Kaneta:2016uyt, Gninenko:2018tlp}. 
The region is summarized as
\begin{align}
2 \times 10^{-4} \leq &~ g'  \leq 2 \times 10^{-3}, 
\label{eq:gp-range} \\
5 \leq &~ m_{Z'} \leq 210~ \mathrm{MeV}
\label{eq:mZp-range}.
\end{align}
The VEV of $\varphi$ is estimated from Eqs.\eqref{eq:gp-range} and \eqref{eq:mZp-range} as 
\begin{align}
v_\varphi = \frac{m_{Z'}}{g'} \simeq 10 - 1000~\mathrm{GeV}.
\end{align}
Since the mass of $\phi$ is roughly given by $\lambda_{\varphi} v_\varphi$, it is naturally expected that $m_\phi$ is the 
same order of $v_\varphi$.
The most stringent bound on the kinetic mixing parameter is set by NA64 \cite{Banerjee:2017hhz}. 
Based on the analysis in \cite{Kaneta:2016uyt}, the constraint from the meson decay is obtained by
\begin{align}
\epsilon \cos\theta_W \sqrt{BR(Z' \rightarrow e^+ e^-)} \leq \epsilon_{\mathrm{MD}},
\end{align}
where $\epsilon_{\mathrm{MD}}$ is the upper bound in \cite{Banerjee:2017hhz}, which depends on $m_{Z'}$. 
For $m_{Z'} = 100~(5)$ MeV, the favored region of $\epsilon$ is obtained as
\begin{align}
\frac{\epsilon}{g'} \leq 2~(0.6),
\end{align}
respectively. 

\begin{figure}[t]
\begin{center}
\includegraphics[width=80mm]{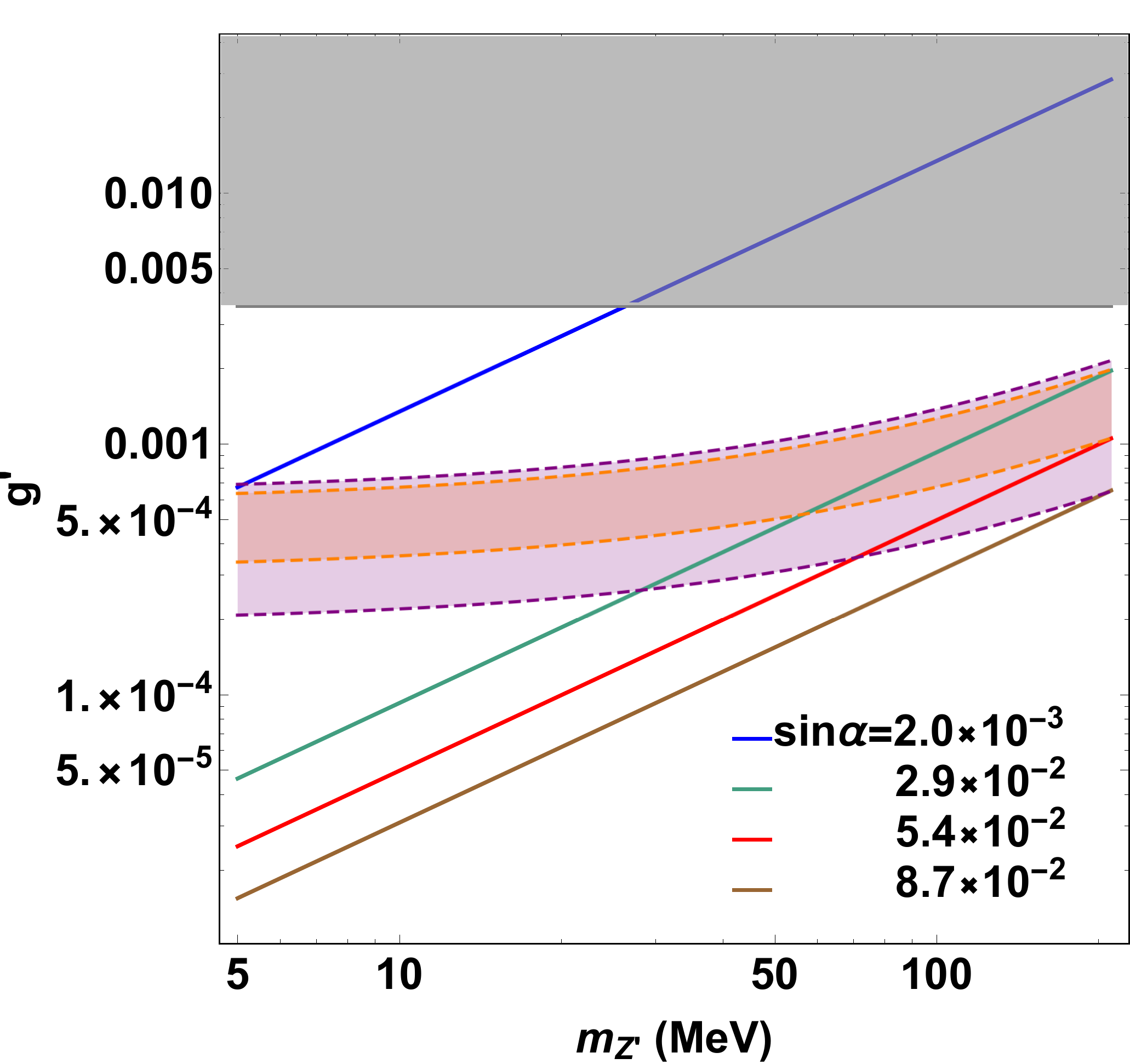} 
~
\includegraphics[width=80mm]{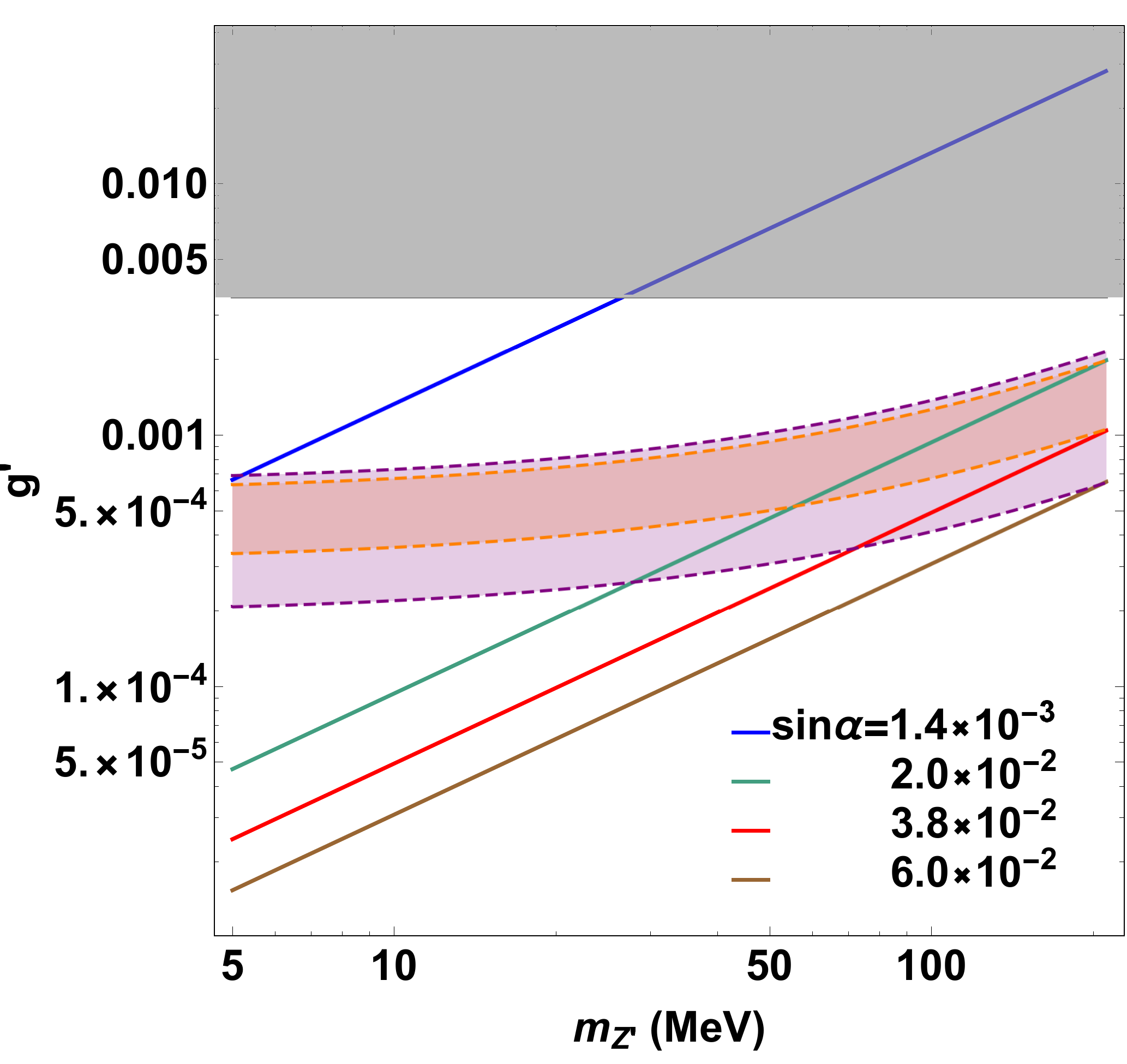} 
\caption{The allowed region of the parameters in $m_{Z'}$-$g'$ plane. In the left and right panels, $m_\phi$ is 
taken as $30$ GeV and larger than $m_{h}/2$, respectively. The blue, green, red and brown lines represents the upper bound on 
the invisible Higgs decays. The values of $\sin \alpha$ corresponding to each line are shown in the figures. 
The favored regions of $(g-2)_\mu$ within $2\sigma$ and $3\sigma$ 
are indicated by the orange and purple bands. The gray area is excluded by the neutrino trident production process.} 
  \label{fig:scalar-mixing}
\end{center}\end{figure}

On the other hand, the scalar mixing and the invisible Higgs decay branching ratio, $BR_{\mathrm{invis}}$, are also constrained by 
analysis of data from the LHC experiment~\cite{Cheung:2015dta,Chpoi:2013wga} as
\begin{align}
\sin \alpha &\leq 0.3,\\
BR_{\mathrm{invis}} &\leq 0.25,
\label{eq:br-inv}
\end{align}

In figure \ref{fig:scalar-mixing}, we show the allowed region of $\sin\alpha$ in $m_{Z'}$-$g'$ plane. In the left and right panels, 
$m_\phi$ is taken as $30$ GeV and larger than $m_{h}/2$, respectively, and $\epsilon/g' = 1$ is assumed.
The blue, green, red and brown lines indicate $BR(h \rightarrow \mathrm{invisibles}) \leq 0.25$ for various values 
of $\sin\alpha$ shown in the figure. The area below lines is allowed. 
The constraint on $g'$ becomes tight as $\sin\alpha$ increases since the decay widths of the invisible Higgs 
decays Eq.\eqref{eq:gamma-Hinvis} are proportional to $\alpha$ when $\alpha \ll 1$.
 The orange and purple regions are the favored region of $(g-2)_\mu$ within $2\sigma$ and $3\sigma$. 
From the right panel, we can see that the scalar mixing, $\sin\alpha$, should be between $1.7\times 10^{-3}$ and $7.5\times 10^{-2}$ 
to explain $(g-2)_\mu$.  This range of $\sin\alpha$ becomes slightly shifted to $2.5\times 10^{-3}$ and $1.1\times 10^{-1}$ 
for $m_\phi \gsim m_h/2$, as shown in the left panel.
Note that for lighter $m_{Z'}$, $\epsilon/g' = 1$ is excluded by NA64. However, when we use $\epsilon/g' = 0.6$, 
the $(g-2)_\mu$ favored region and the excluded region are slightly shifted upward in this case. 
Therefore, the result does not change so much.
In the following analysis, we fix $m_{Z'} = 100$ MeV, $\epsilon/g' = 1$  
and $\sin\alpha = 0.05$, and discuss the observation possibilities at the LHC and the ILC collider experiments.

\section{Signature of extra scalar boson and $Z'$ in collider experiments}

In this section, we discuss signature of $\phi$ and $Z'$ in collider experiments;  the LHC 
and the ILC.
We consider the mass of $\phi$ and $Z'$ are $O(10-100)$ GeV and $O(100)$ MeV, respectively.
The scalar boson $\phi$ can be produced in collider experiments through the mixing with 
the SM Higgs boson, and dominantly decays into $Z'$ bosons. 
As we showed in the previous section, $Z'$ dominantly decays into $\nu \bar{\nu}$, 
and subdominantly into $e^+ e^-$ for $\epsilon/g' < 1$. 
We investigate possibilities to search for the signature of $\phi$ and $Z'$ in collider experiments in this situation.

\subsection{Signatures at the LHC}

In the parameter space of our choice, the gauge boson $Z'$ is mainly produced from $\phi$ decay at the LHC 
because $Z'$ interacts with quarks only through the kinetic mixing.  The main production of $\phi$ is gluon fusion through the mixing with the SM Higgs.

To identify the gauge and scalar bosons, $Z'$ should decay into $e^+e^-$ because $Z'$ and $\phi$ are electrically neutral. 
However, $e^+e^-$ pair from $Z'$ decay will be highly collimated due to lighter $Z'$ mass  than GeV scale. Here we estimate the degree of collimation; 
if $Z' \to e^+ e^-$ decay system is boosted with velocity of $v_{Z'} \sim \sqrt{m_\phi^2/4-m_{Z'}^2}/(m_\phi/2)$ 
which is induced by decay of $\phi \to Z' Z'$, 
the angle between $e^+$ and $e^-$ is approximately $\theta \sim \cos^{-1} (1-8m_{Z'}^2/m_\phi^2)$ where we assumed $e^\pm$ direction before boost is $z$-direction and $\vec{v}_{Z'}$ is perpendicular to the direction.
Then the angle is $\sim 1^\circ$ for $m_{Z'} = 100$ MeV and $m_\phi = 50$ GeV.
It is discussed in \cite{Aad:2015sva, ATLAS:2017lvz} that reconstruction of such a collimated $e^+e^-$ pair is 
experimentally challenging due to angle resolution with the ATLAS detector. 
The reconstruction of $e^+ e^-$ pair is possible for $m_{Z'} \geq 15$ GeV, which is already excluded for muon $(g-2)$ to be 
explained. Even for $\mu^+ \mu^-$ pair, the reconstruction has been simulated only above $m_{Z'} \geq 1$ GeV. 
A new analysis would be needed for the reconstruction of $e^+$ and $e^-$ momenta. However such a new analysis is 
beyond the scope of this paper and we do not discuss here. From this fact,  lepton colliders are 
more suitable to search for $\phi$ 
in our parameter choice because it can use missing energy search due to the precise knowledge of the initial energy. 

\begin{figure}[t]
\begin{center}
\includegraphics[width=70mm]{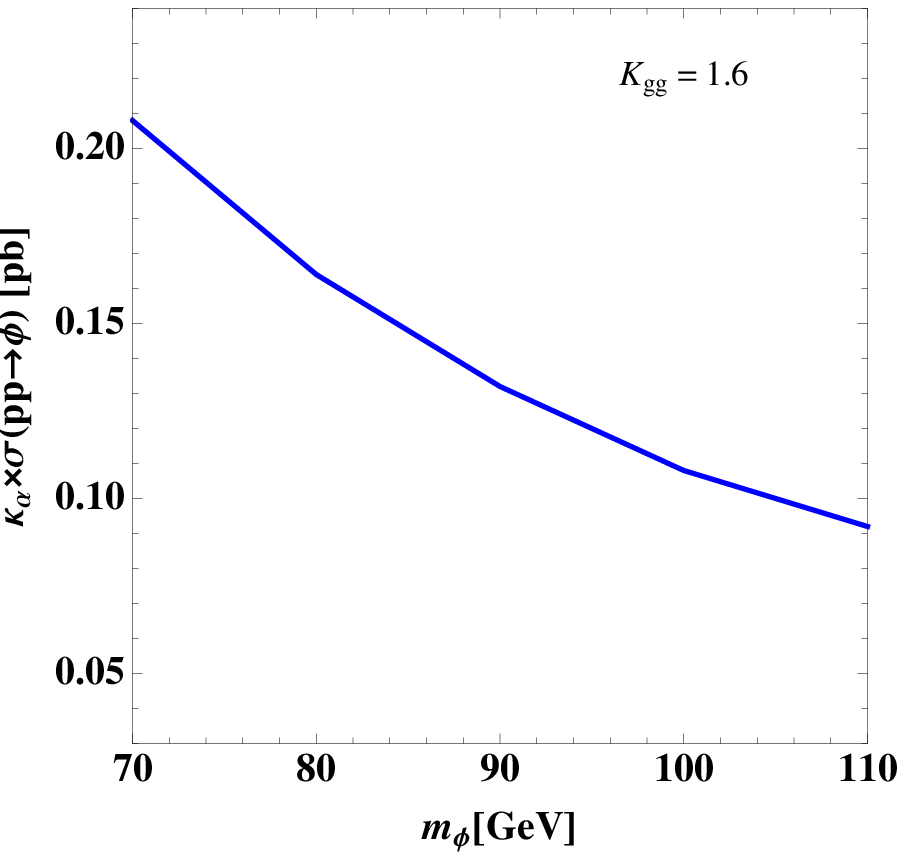} \quad
\includegraphics[width=70mm]{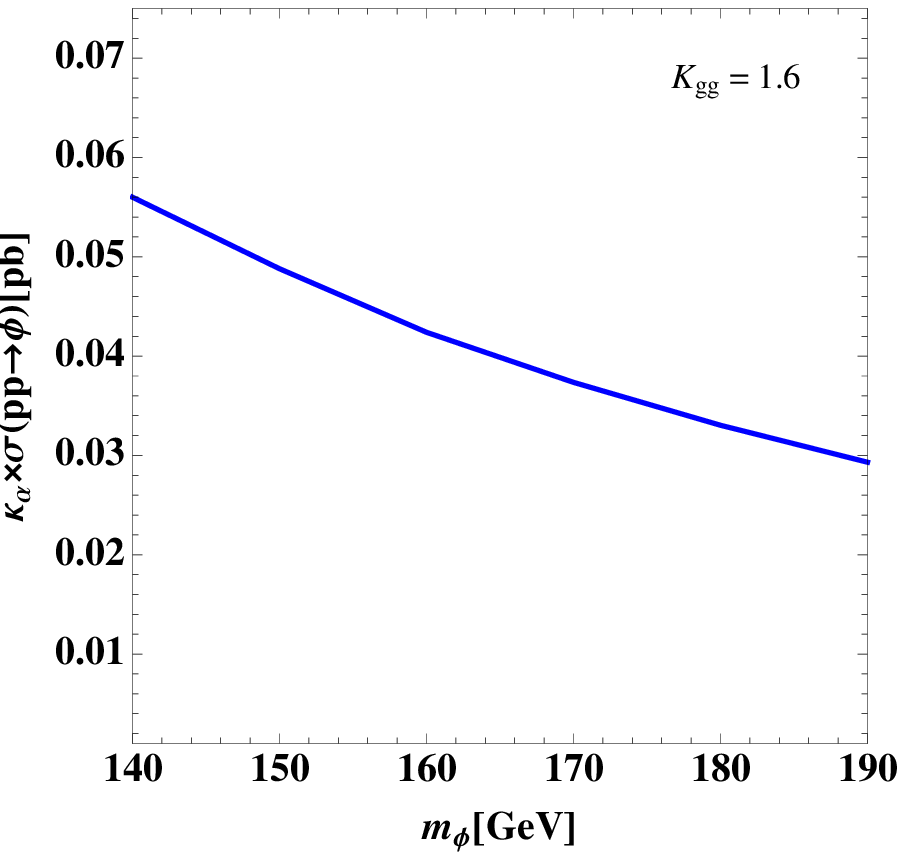}
\caption{The cross section for $pp \to \phi$ as a function of $m_\phi$ which is multiplied by scaling factor $\kappa_\alpha = (0.05/\sin \alpha)^2$, and $\sqrt{s} = 13$ TeV is applied.} 
  \label{fig:CSphi}
\end{center}\end{figure}

Although the light $Z'$ is hard to observe at the LHC, for future reference, we show the production cross section of $\phi$ via gluon fusion 
process $gg \to \phi$ through the mixing with the SM Higgs boson.
The relevant effective interaction for the gluon fusion is given by~\cite{Gunion:1989we}
\begin{equation}
{\mathcal L}_{\phi gg} = \frac{\alpha_s}{16 \pi} \frac{\sin \alpha}{v} A_{1/2}(\tau_t) \phi G^a_{\mu \nu} G^{a \mu \nu}, 
 \end{equation}
 where $G^a_{\mu \nu}$ is the field strength for gluon and $A_{1/2}(\tau_t) = -\frac{1}{4} [\ln[(1+\sqrt{\tau_t})/(1-\sqrt{\tau_t})] - i \pi ]^2$ with $\tau_t = 4 m_{t}^2/m_\phi^2$.
 This effective interaction is induced from $\bar t t \phi$ coupling via the mixing effect where we take into account only top Yukawa coupling since the other contributions are subdominant.
In Fig.~\ref{fig:CSphi}, we show the production cross section estimated by {\tt MADGRAPH5}~\cite{Alwall:2014hca} implementing the effective interaction by use of FeynRules 2.0 \cite{Alloul:2013bka}, which is multiplied by scaling factor $\kappa_\alpha \equiv (0.05 /\sin \alpha)^2$ since the cross section is proportional to $\sin^2 \alpha$.
We also included K-factor of $K_{gg} = 1.6$ for gluon fusion process which comes from NLO correction~\cite{Djouadi:2005gi}. 
We can see that the production cross section is $\mathcal{O}(0.01 - 0.1)$ pb for $70 \leq m_\phi \leq 190$ GeV. 
Assuming the integrated luminosity $300$ fb$^{-1}$ (LHC) and $3000$ fb$^{-1}$ (HL-LHC), the number of $\phi$ produced is $\mathcal{O}(10^3-10^4)$ and $\mathcal{O}(10^4-10^5)$, respectively. Therefore we have sizable number of events, and 
background estimation as well as analysis on the collimated $e^+ e^-$ pair will be important.

\subsection{Signatures at the ILC}

Here we discuss $\phi$ production processes and possibility to search for its signature at  the ILC experiment. 
In lepton collider experiments, $\phi$ can be produced by the processes such that $e^+ e^- \to Z \phi$, $e^+ e^- \to \nu \bar \nu \phi$ and $e^+ e^- \to e^+ e^- \phi$
where the second process is $W$ boson fusion and the third process is $Z$ boson fusion; these processes are induced by the interactions in Eq.~(\ref{eq:intV}).
Remarkably, polarized electron and positron beams will be available at ILC where possible combinations 
of $(e^+, e^-)$ polarization is $(-+, +-, ++, --)$. 
In our following analysis, we apply fractions of $(45 \%, 45 \%, 5 \%, 5 \%)$ with the total integrated luminosity $L = 2000$ fb$^{-1}$, and $\pm 80(30) \%$ polarization for the electron(positron) beam as a realistic value~\cite{Fujii:2017vwa}. 
To simplify the analysis, we only consider $(e^+, e^-)$ polarizations $(+,-)$ and $(-,+)$ with the 
integrated luminosity of $900$ fb$^{-1}$ where we respectively denote these cases as $LL$ and $RR$ polarizations hereafter.
\begin{figure}[t]
\begin{center}
\includegraphics[width=70mm]{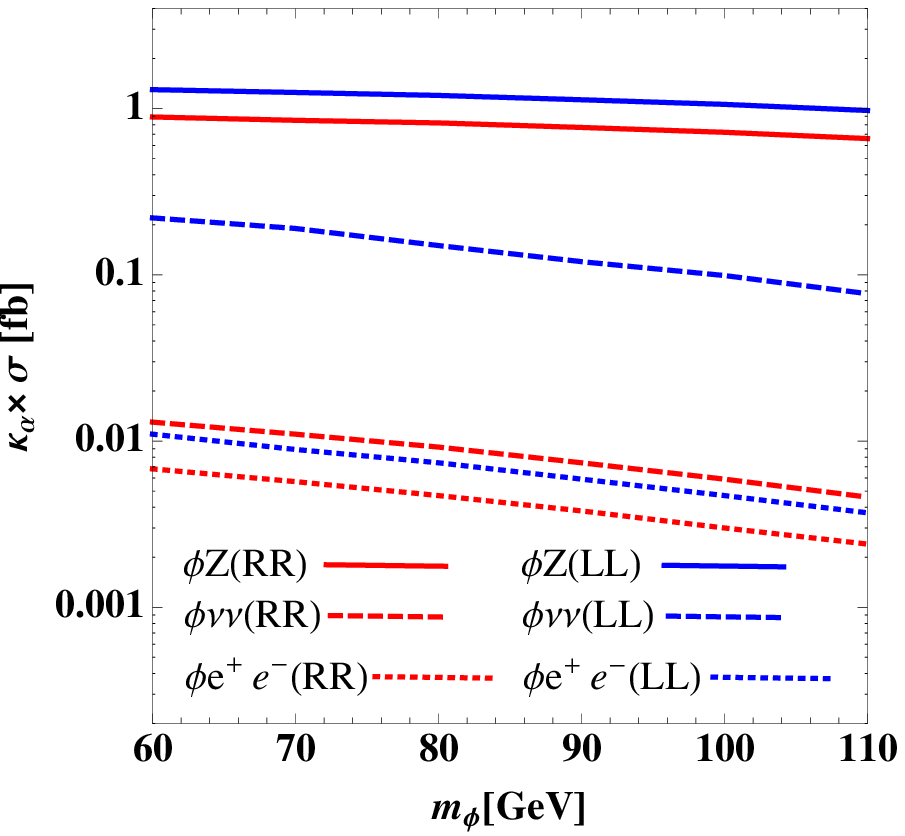} \quad
\includegraphics[width=70mm]{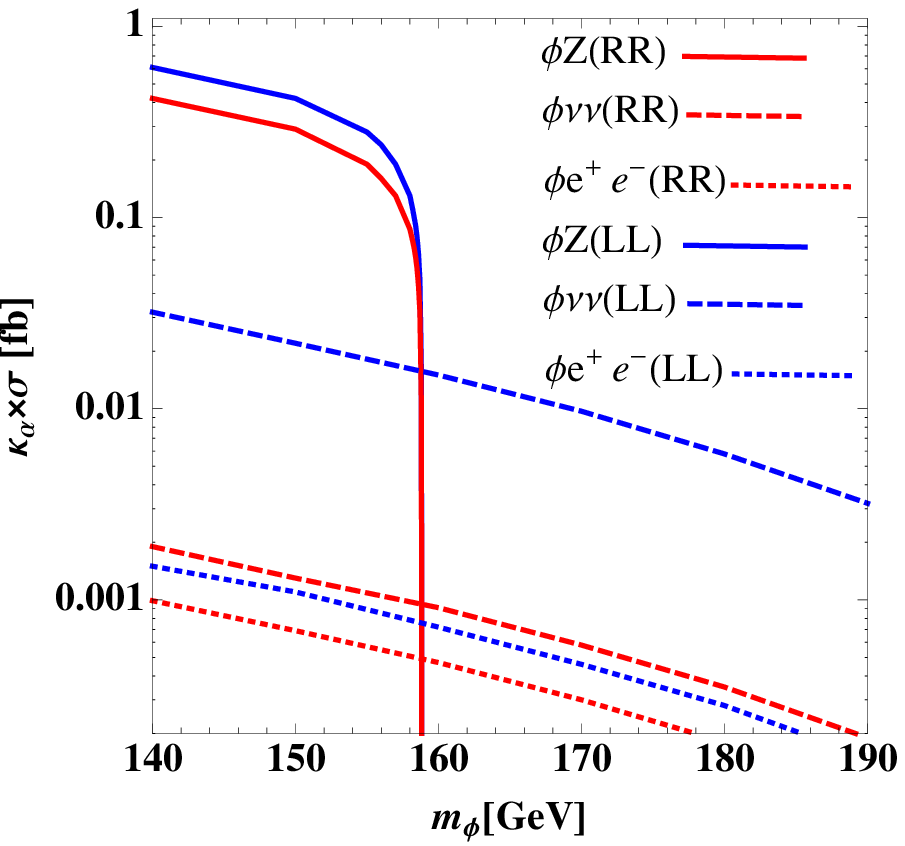}
\caption{The cross section for $\phi$ production processes in $e^+e^-$ collider for two polarization cases $LL$ and $RR$ as a function of $m_\phi$. The scaling factor $\kappa_\alpha = (0.05/\sin \alpha)^2$ is multiplied with the cross section. } 
  \label{fig:CSphiEC}
\end{center}\end{figure}
In Fig.~\ref{fig:CSphiEC}, we show the production cross sections for $\sqrt{s}=250$ GeV for two polarization cases calculated by {\tt CalcHEP 3.6}~\cite{Belyaev:2012qa} implementing relevant interactions, which is scaled by $\kappa_\alpha = (0.05/\sin \alpha)^2$ factor. The figure shows that $e^+ e^- \to Z \phi$ mode gives the largest cross section for $m_\phi \lesssim 160$ GeV 
for the $LL$ and $RR$ polarizations. In our following analysis, we thus focus on the $Z \phi$ mode since cross sections for the other modes are small.
Then we consider two cases; (1) $Z$ decays into two leptons, $\ell^+ \ell^-$ ($\ell = e, \mu$) and (2) $Z$ decays into two jets, $jj$. 
In both cases, $\phi$ decays as $\phi \to Z' Z' \to \nu \nu \bar \nu \bar \nu$ which is the dominant decay mode. Therefore our signals are
\begin{equation} 
\ell^+ \ell^- + \slashed{E},  \quad jj + \slashed{E}
\end{equation}
for cases (1) and (2) respectively where $\slashed{E}$ denotes missing energy. Note that we can reconstruct mass of $\phi$ in lepton collider experiments using energy momentum conservation even if $\phi$ becomes missing energy.

Hereafter we perform a simulation study of our signal and background (BG) processes in both cases (1) and (2); the events are generated via {\tt MADGRAPH/MADEVENT\,5}~\cite{Alwall:2014hca}, where the necessary Feynman rules and relevant parameters of the model are implemented by use of FeynRules 2.0 \cite{Alloul:2013bka},
the {\tt PYTHIA\,6}~\cite{Ref:Pythia}  is applied to deal with hadronization effects,  the  initial-state radiation (ISR) and final-state radiation (FSR) effects and the decays of the SM particles, and {\tt Delphes}~\cite{Delphes} is used for detector level simulation.

\subsubsection{The case of $\ell^+ \ell^- \slashed{E}$ signal}

Here we discuss the "$\ell^+ \ell^- + \slashed{E}$" signal and corresponding BG events. We then estimate 
the discovery significance applying relevant kinematical cuts.
In this case we consider following BG processes:
 \begin{itemize}
 \item $e^+ e^- \to \ell^+ \ell^- \nu \bar \nu$ \,, 
 \item $e^+ e^- \to \tau^+ \tau^+$,
\end{itemize} 
where the first process mainly comes from $e^+e^- \to ZZ/W^+W^-$ followed by leptonic decays of $Z/W^\pm$ 
while the second process gives $\ell^+ \ell^- + \slashed{E}$ evens via leptonic decay of $\tau^\pm$.
Signal and BG events are generated with basic cuts implemented in {\tt MADGRAPH/MADEVENT\,5} as 
\begin{equation}
\label{eq:BasicCutsL}
p_T(\ell^\pm) > 7 \ {\rm GeV}, \quad |\eta(\ell^\pm)| < 2.5,
\end{equation} 
where $p_T$ denotes transverse momentum and $\eta = -\ln (\tan \theta/2)$ is the pseudo-rapidity with $\theta$ being the scattering angle in the laboratory frame.
With the basic cuts, the cross sections for the BG processes are obtained such as 
\begin{align}
& \sigma (e^+ e^- \to \ell^+ \ell^- \nu \bar \nu) = 1.99 (0.186) \ {\rm pb} \quad \text{for LL(RR) polarization}, \\
& \sigma (e^+ e^- \to \tau^+ \tau^+) = 2.36 (1.94) \ {\rm pb} \quad \text{for LL(RR) polarization}, \label{eq:BGtau}
\end{align}
where detector efficiency is not applied here.
Note that $\ell^+ \ell^- \nu \bar \nu$ background is small for $RR$ polarization since $W^+ W^-$ production cross section is suppressed.

\begin{figure}[t]
\begin{center}
\includegraphics[width=50mm]{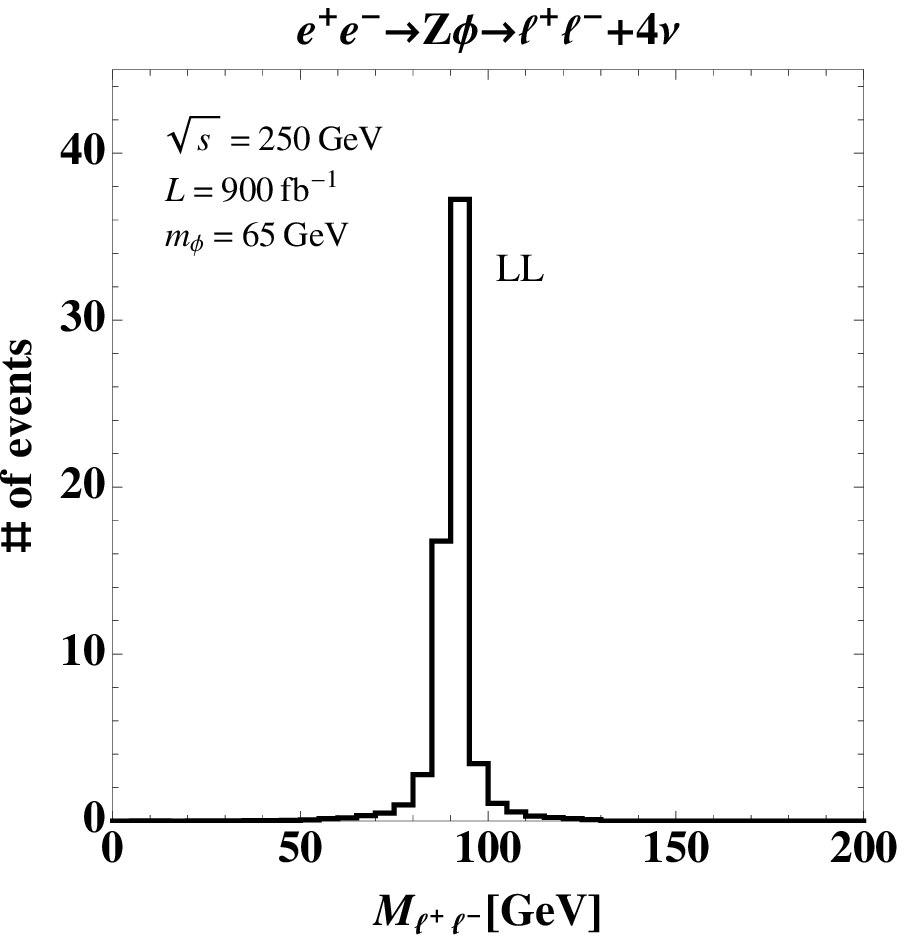} \quad
\includegraphics[width=50mm]{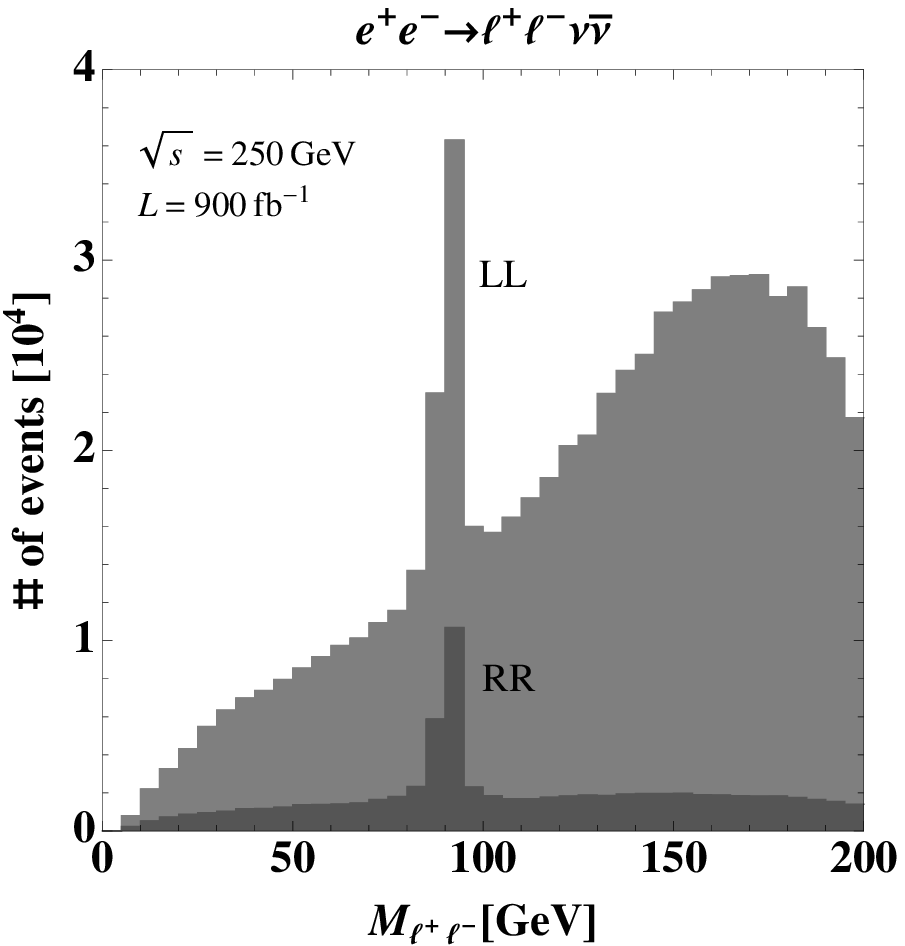} \quad
\includegraphics[width=50mm]{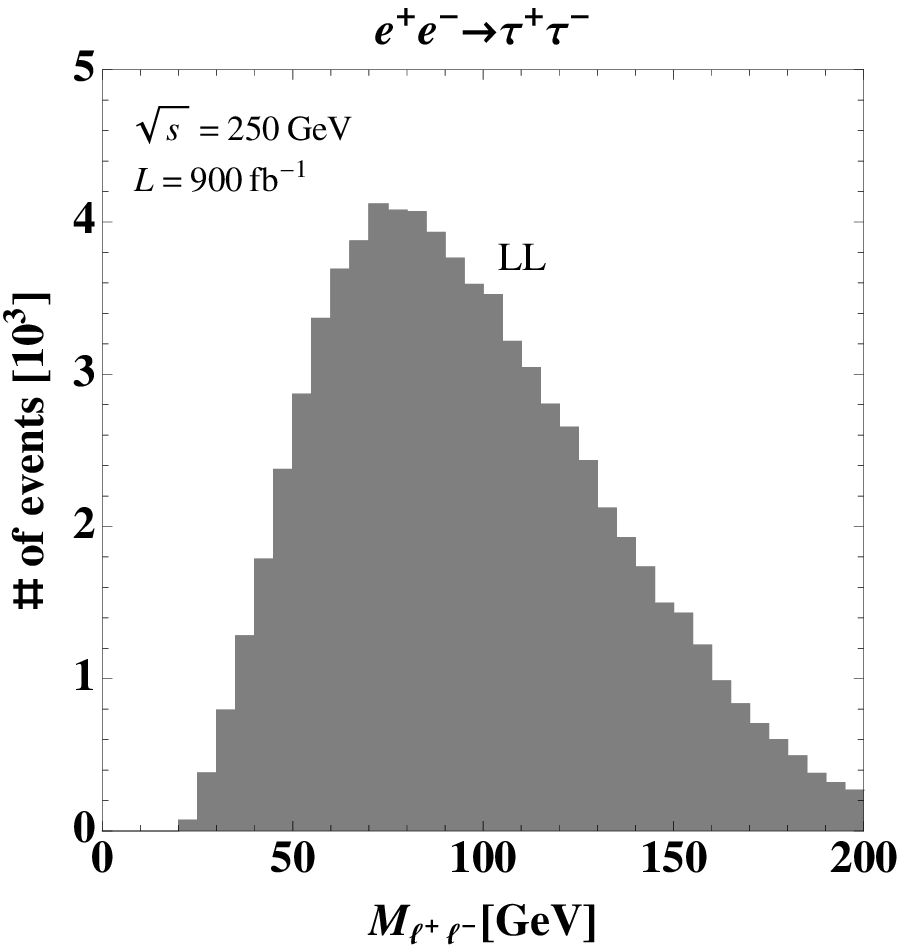} 
\caption{Distribution of invariant mass for $\ell^+ \ell^-$ with only basic cuts where left-, middle- and right-panels correspond to signal, $\ell^+\ell^- \nu \bar \nu$ BG and $\tau^+ \tau^-$ BG events.  Here $\kappa_\alpha =1$ is applied.} 
  \label{fig:MeeDist}
\end{center}\end{figure}
\begin{figure}[t]
\begin{center}
\includegraphics[width=50mm]{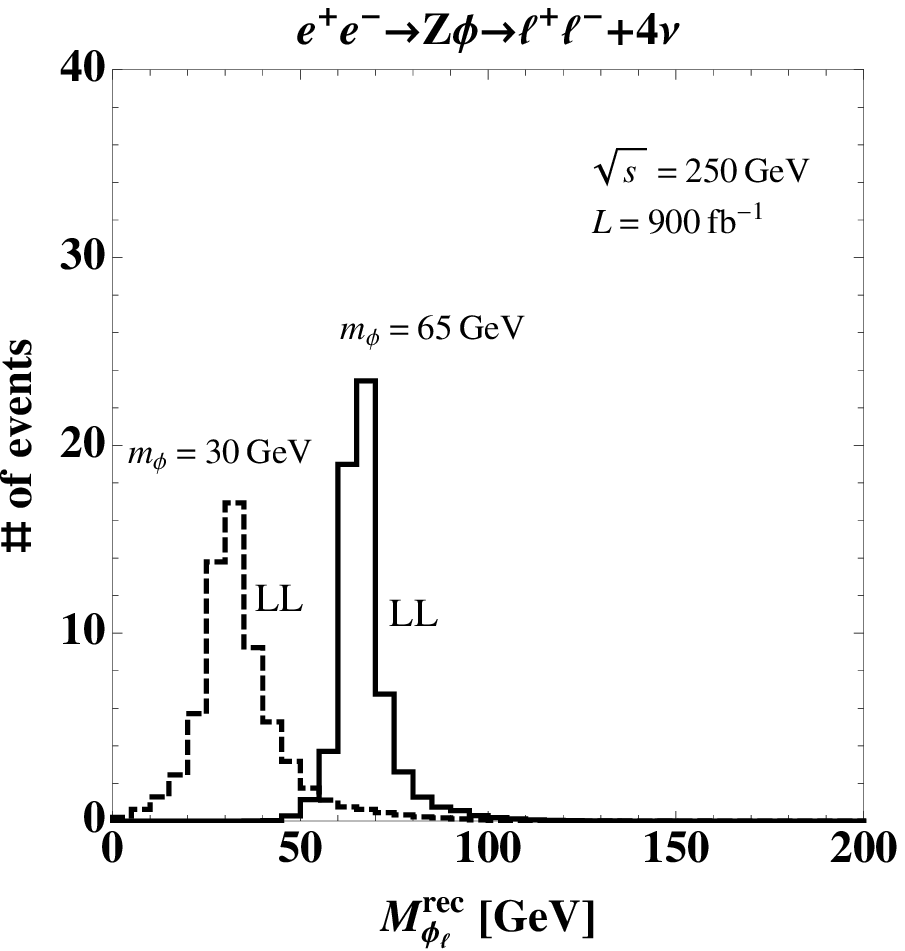} \quad
\includegraphics[width=50mm]{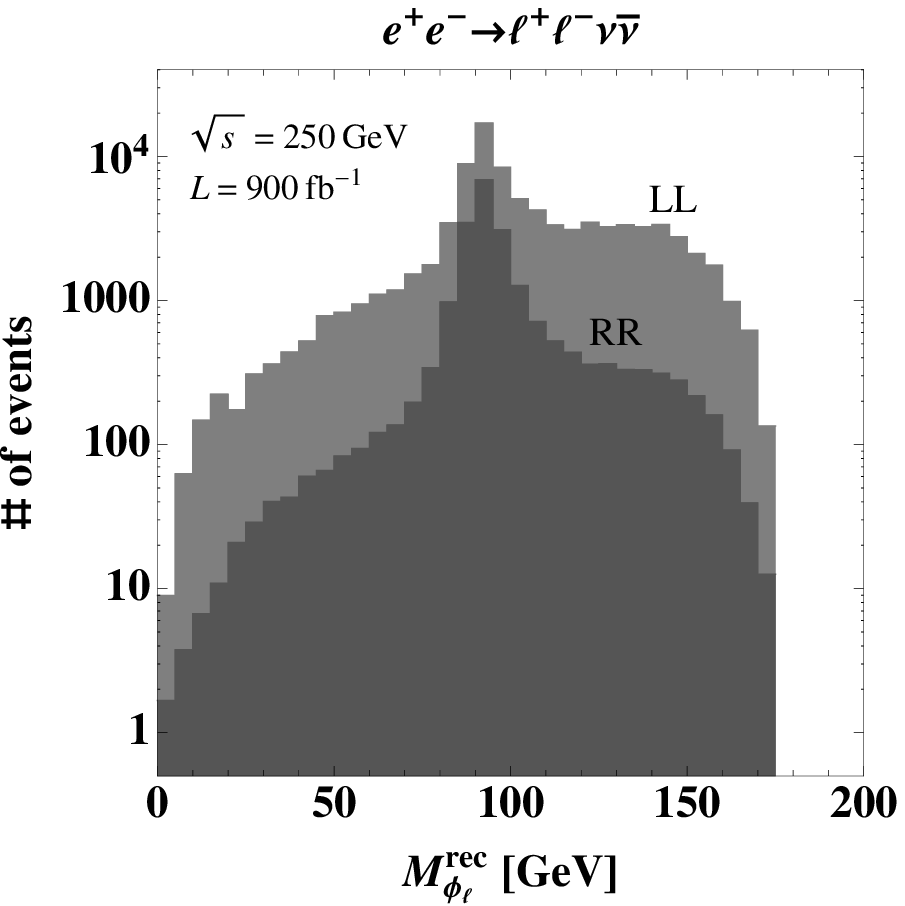} \quad
\includegraphics[width=50mm]{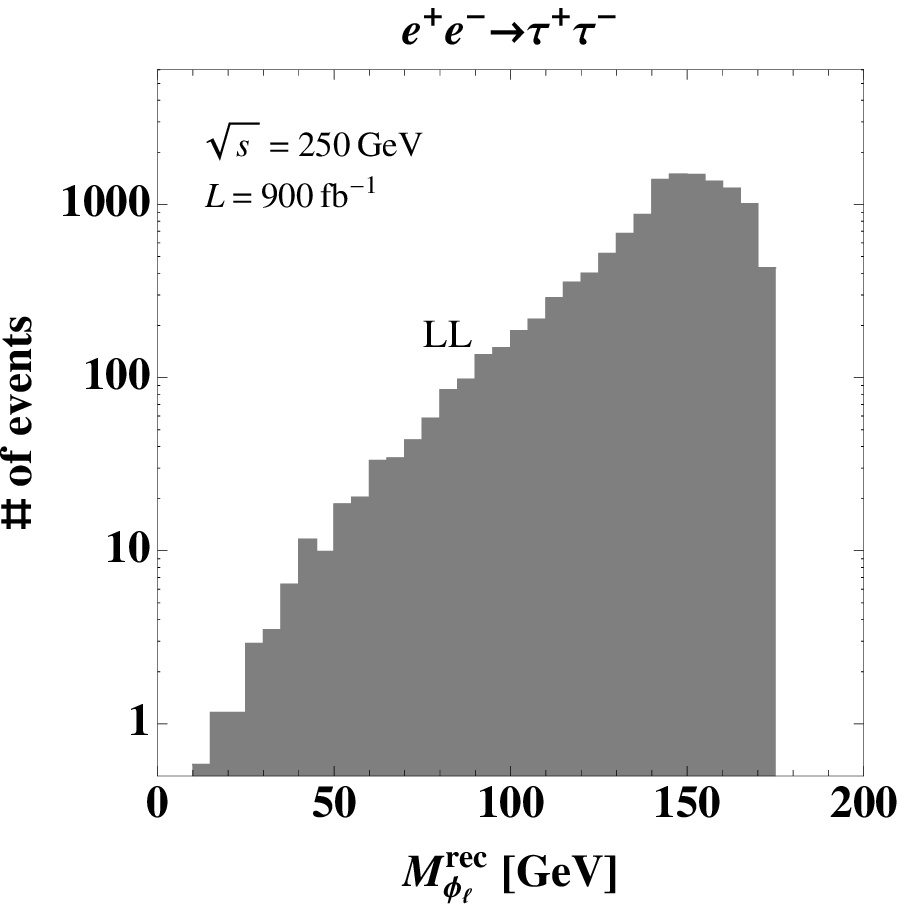} 
\caption{Distribution of reconstructed $\varphi$ mass after imposing basic and $M_{\ell^+ \ell^-}$ cuts where left-, middle- and right-panels correspond to signal, $\ell^+\ell^- \nu \bar \nu$ BG and $\tau^+ \tau^-$ BG events. Here $\kappa_\alpha =1$ is applied.} 
  \label{fig:MphiDist}
\end{center}\end{figure}

We then investigate kinematic distributions for signals and BGs,  and also efficiency of kinematical cutoff.
Plots in Fig.~\ref{fig:MeeDist} show $\ell^+ \ell^-$ invariant mass distributions where left-, middle- and right-panels correspond to events from the signal, $\ell^+\ell^- \nu \bar \nu$ BG and $\tau^+ \tau^-$ BG with only basic cuts in Eq.~(\ref{eq:BasicCutsL}).
Here we show distribution for both $LL$ and $RR$ polarizations in $\ell^+ \ell^- \nu \bar \nu$ BG, 
and those for only $LL$ polarization are shown in the other plots since $RR$ case present almost the same behavior.
We find that the distribution for signal events shows a clear peak at $Z$ boson mass. On the other hand the distribution for $\ell^+ \ell^- \nu \bar \nu$ BG has a peak at $Z$ mass and continuous region coming from $e^+e^- \to W^+W^-$ process. 
Note that continuous region is much suppressed in $RR$ case since contribution from $e^+ e^- \to W^+ W^-$ is small. 
The distribution for $\tau^+ \tau^-$ BG has broad bump peaked around 80 GeV.
To reduce the BG events, we thus impose the $\ell^+ \ell^-$ invariant mass cuts as 
\begin{equation}
m_Z - 10 \ {\rm GeV} < M_{\ell^+ \ell^-} < m_Z + 10 \ {\rm GeV}.
\end{equation}
Furthermore we reconstruct the mass of $\phi$ using energy momentum conservation. 
The reconstructed mass is given by
\begin{equation}
M_{\phi_\ell}^{rec} = \sqrt{s + m_Z^2 -2 (E_{\ell^+} + E_{\ell^-}) \sqrt{s} }
\end{equation}
where $E_{\ell^\pm}$ is energy of final state $\ell^\pm$.
Plots in Fig.~\ref{fig:MphiDist} show the distribution of $M_\phi^{rec}$ for the signal and BGs.
As in the $M_{\ell^+ \ell^-}$ distribution, we show the distribution for both $LL$ and $RR$ polarizations in $\ell^+ \ell^- \nu \bar \nu$ BG and show only those for $RR$ polarization in the other plots.
We see that the mass of $\phi$ is indeed reconstructed giving clear peaks. 
Note also that $\ell^+ \ell^- \nu \bar \nu$ BG has a peak at $Z$ boson mass which comes from $e^+ e^- \to ZZ$ process due to energy momentum conservation.
Then we also impose kinematical cuts for $M_\phi^{rec}$ such that
\begin{equation}
m_\phi - 10 \ {\rm GeV} < M_{\phi_\ell}^{rec} < m_\phi + 10 \ {\rm GeV}.
\end{equation}
Table~\ref{tab:cut_effectR} summarizes the effect of kinematical cuts to signal and BGs for $RR$ polarization as an example where cut efficiency has similar behavior in $LL$ polarization.
We see that the number of events for the BGs can be highly reduced by the $M_{\ell^+ \ell^-}$ and $M_{\phi_\ell}^{rec}$ cuts while that of the signal events does not change significantly.
Note that the number of the BG events is large in the region $M_{\phi_\ell}^{rec} \gtrsim 80$ GeV. It would be difficult to search for our signal if $m_\phi$ is in the region.

Finally we estimate the discovery significance by
\begin{equation} 
\label{eq:sig}
S_{cl} = \frac{N_S}{\sqrt{N_{BG}}},
\end{equation}
where $N_S$ and $N_{BG}$ respectively denote the number of events for the signal and total BG.
The significances before and after kinematical cuts are shown in the last column of Table~\ref{tab:cut_effectR} for $RR$ polarization.
We see that cut for $M_{\phi_\ell}^{rec}$ can reduce the BG events significantly while keeping signal events.
In addition, we compare the significances in $RR$ and $LL$ polarizations, and sum of them after all kinematical cuts in Table.~\ref{tab:cut_effectLR}.
Then we find that the events from only $RR$ polarization provides the largest significance since $\ell^+ \ell^- \nu \bar \nu$ background in $LL$ polarization is large and hence decrease the significance.
We can obtain discovery significance of 2.2(2.5) for $m_\phi = 65(30)$ GeV with $\kappa_\alpha = 1$ corresponding to $\sin \alpha = 0.05$ in $RR$ polarization.
Thus small scalar mixing as $\sin \alpha =0.05$ will be constrained when mass of $\phi$ is as light as 65 GeV for $RR$ polarization. 
Furthermore if $\sin \alpha \sim 0.1$ we can get discovery significance larger than $S_{cl} = 5$ since $\kappa_\alpha \sim 1/4$. 
Note that more detailed kinematical cuts will improve the significance~\cite{Drechsel:2018mgd} but it is beyond the scope of this paper. 

\begin{center} 
\begin{table}
\begin{tabular}{|c | c| c| c |c |}\hline
& $\kappa_\alpha N_{S}^{\kappa_\alpha=1}$; $m_\phi = (65, 30)$ GeV & \quad $N_{BG}^{\ell^+\ell^-\nu \bar \nu}$  \quad & \quad $N_{BG}^{\tau\tau}$ \quad & \quad $\kappa_\alpha S_{cl}^{\kappa_\alpha=1}$ \quad \\ \hline
Only basic cuts & ($51.$, $53.$)  & $7.7 \times 10^4$ & $6.3 \times 10^4$  &  (0.14, 0.14) \\ 
+ $M_{\ell^+ \ell^-}$ cut & ($48.$, $49.$) &  $2.1 \times 10^4$ & $1.3 \times 10^4$  & (0.25, 0.27)   \\ \hline \hline
+ $M_{\phi_\ell}^{rec}$ cut for $m_\phi=65$ GeV & ($42.$, $\cdots$) & $2.2 \times 10^2$ & $1.3 \times 10^2$ & (2.2, $\cdots$) \\ 
+ $M_{\phi_\ell}^{rec}$ cut for $m_\phi=30$ GeV & ($\cdots$, $34.$) & $1.7 \times 10^2$ & $14.$ & ($\cdots$, 2.5) \\ \hline
\end{tabular}
\caption{The number of events for signal ($N_S$), BG ($N_{BG}$) and significance ($S_{cl}$) for $RR$ polarization case after each cut where we have adopted $m_\phi= (30, 65)$ GeV as reference values. The integrated luminosity is taken as $900$ fb$^{-1}$, and  $N_S(S_{cl})$ is given by the products of scaling factor $k_\alpha$ and the value for $\kappa_\alpha =1$.}
\label{tab:cut_effectR}
\end{table}
\end{center}

\begin{center} 
\begin{table}
\begin{tabular}{|c | c| c| c |c |}\hline
& $\kappa_\alpha N_{S}^{\kappa_\alpha=1}$; $m_\phi = 65(30)$ GeV & \quad $N_{BG}^{\ell^+\ell^-\nu \bar \nu}$  \quad & \quad $N_{BG}^{\tau\tau}$ \quad & \quad $\kappa_\alpha S_{cl}^{\kappa_\alpha=1}$ \quad \\ \hline
$RR$ & $42.(34.)$  & $2.2(1.7) \times 10^2$ & $1.3(0.14)\times 10^2$  &  $2.2(2.5)$  \\ 
 $LL$ & $53.(47.)$ &  $4.7(1.7) \times 10^3$ & $1.6(0.15) \times 10^2 $  & $0.75(1.1)$   \\ \hline \hline
$LL + RR$ & $95.(81.)$ & $4.9(1.9) \times 10^3$ & $2.9(0.29) \times 10^2$ & $1.3(1.8)$ \\  \hline
\end{tabular}
\caption{The number of events for signal ($N_S$), BG ($N_{BG}$) and significance ($S_{cl}$) for $RR$ and $LL$ polarizations with integrated luminosity of $900$ fb$^{-1}$ each and for sum of events from two polarizaitons, where we show cases for $m_{\phi} = 65(30)$ GeV with all kinematical cuts imposed.}
\label{tab:cut_effectLR}
\end{table}
\end{center}

\subsubsection{The case of $j j + \slashed{E}$ signal}

Here we discuss the "$j j + \slashed{E}$" signal and corresponding BG events and estimate discovery significance applying relevant kinematical cuts.
In this case we consider following BG processes:
 \begin{itemize}
 \item $e^+ e^- \to j j \nu \bar \nu$ \,, 
 \item $e^+ e^- \to \tau^+ \tau^+$,
\end{itemize} 
where the first process mainly comes from $e^+e^- \to ZZ$ followed by $Z$ decay into jets/neutrinos and the second process gives $j j + \slashed{E}$ events due to miss-identification of $\tau$-jet as hadronic jets with missing energy.
Signal and BG events are generated with basic cuts for jets in final states implemented in {\tt MADGRAPH/MADEVENT\,5} as 
\begin{equation}
\label{eq:BasicCutsJ}
p_T(j) > 20 \ {\rm GeV}, \quad \eta(j) < 5.0 \, .
\end{equation}
With the basic cuts, the cross sections for BG processes are obtained such as 
\begin{align}
& \sigma (e^+ e^- \to j j \nu \bar \nu) = 0.398 (0.158) \ {\rm pb} \quad \text{for LL(RR) polarization}, 
\end{align}
where efficiency at the detector is not applied here and cross section for $\tau^+ \tau^-$ is the same as Eq.~(\ref{eq:BGtau}).

\begin{figure}[t]
\begin{center}
\includegraphics[width=50mm]{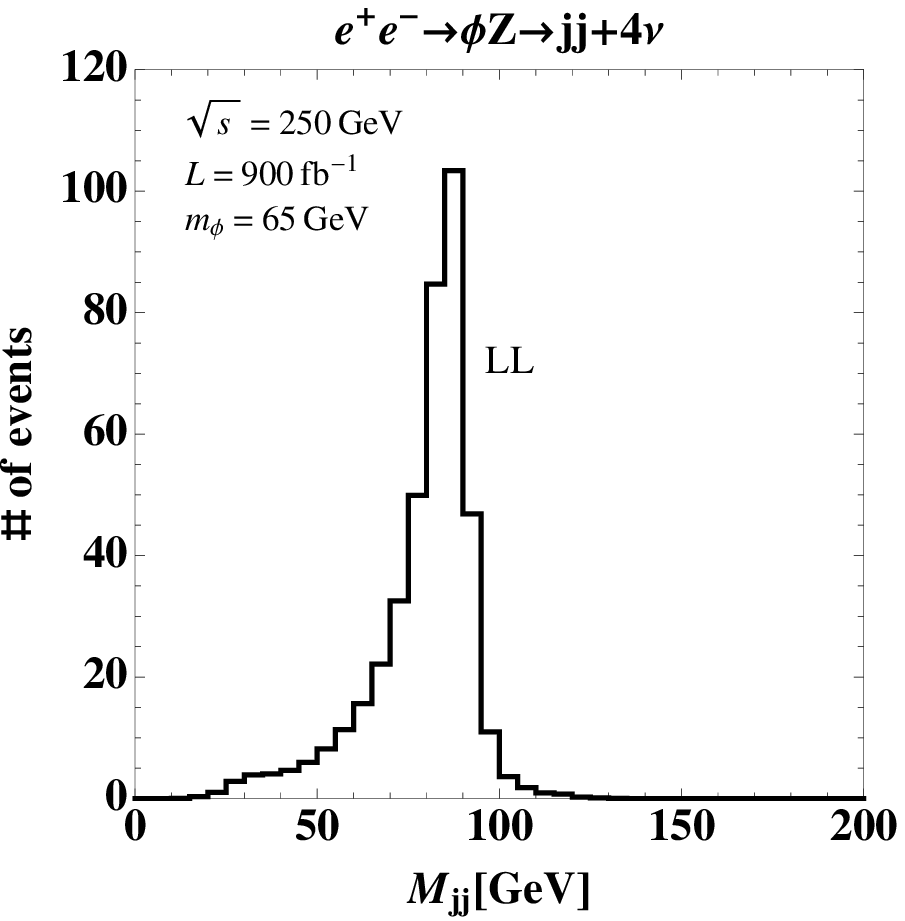} \quad
\includegraphics[width=50mm]{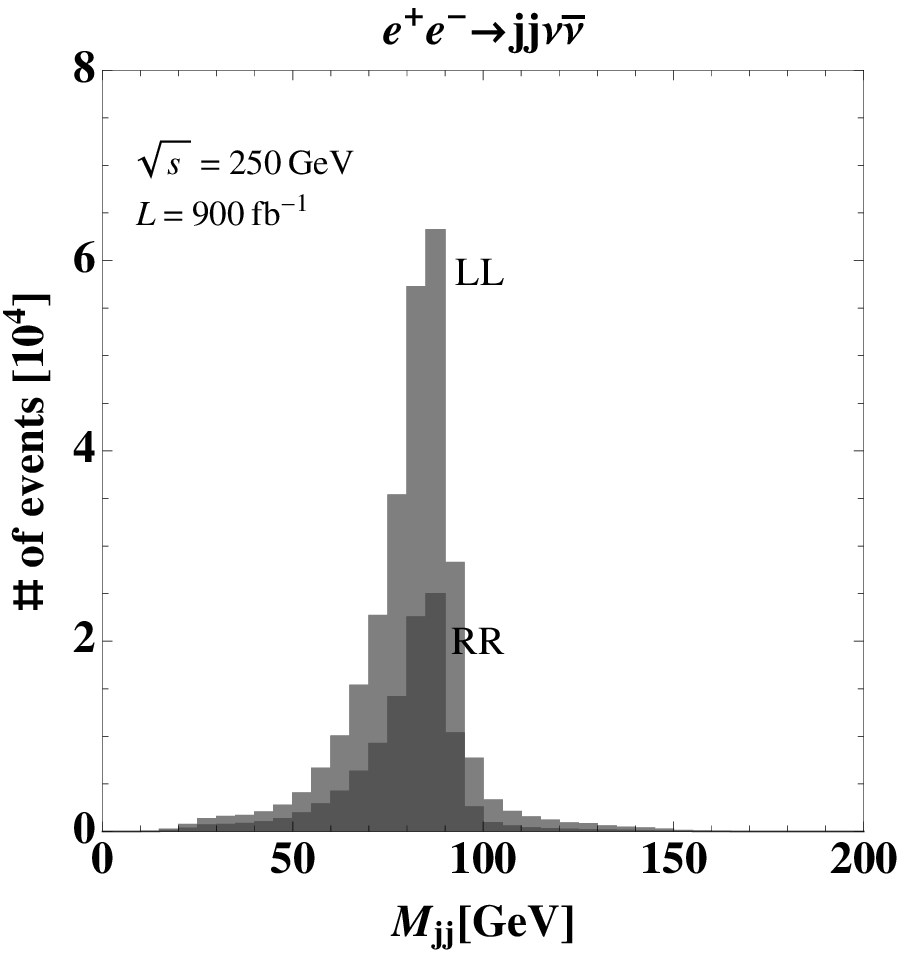} \quad
\includegraphics[width=50mm]{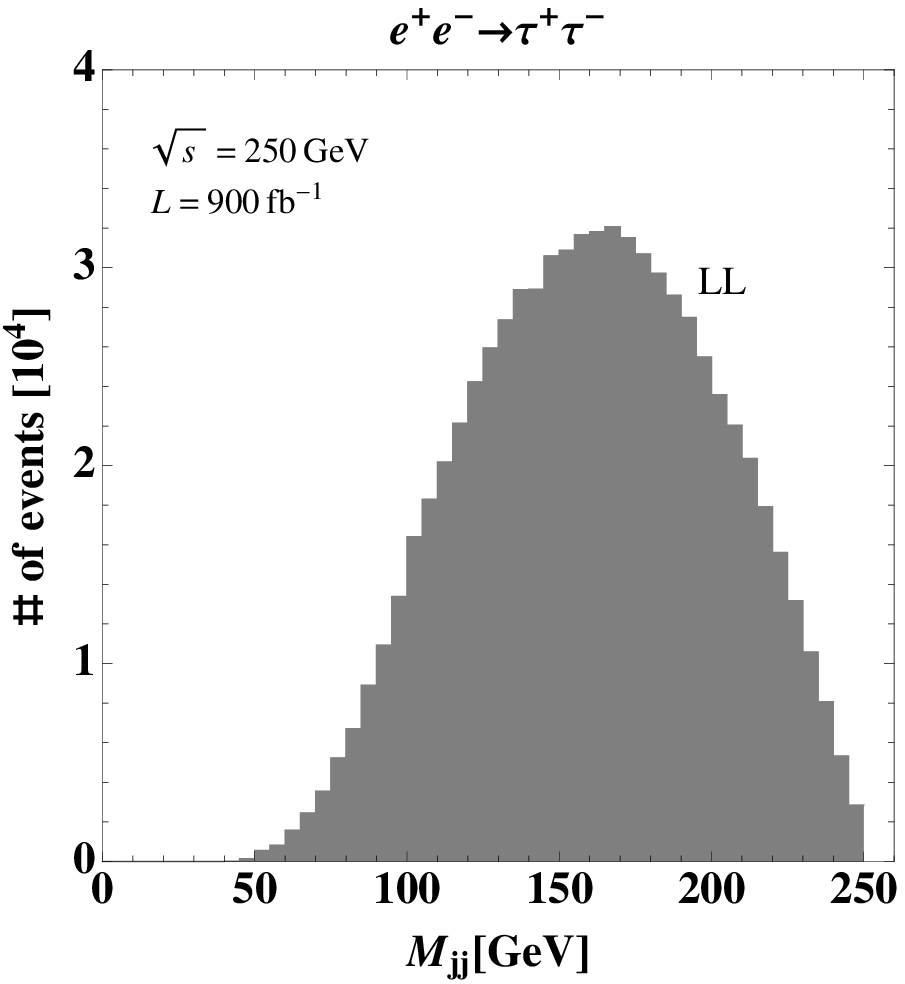} 
\caption{Distribution of invariant mass for two jets with only basic cuts where left-, middle- and right-panels correspond to signal, $jj \nu \bar \nu$ BG and $\tau^+ \tau^-$ BG events.  Here $\kappa_\alpha =1$ is applied.} 
  \label{fig:MjjDist}
\end{center}\end{figure}
\begin{figure}[t]
\begin{center}
\includegraphics[width=50mm]{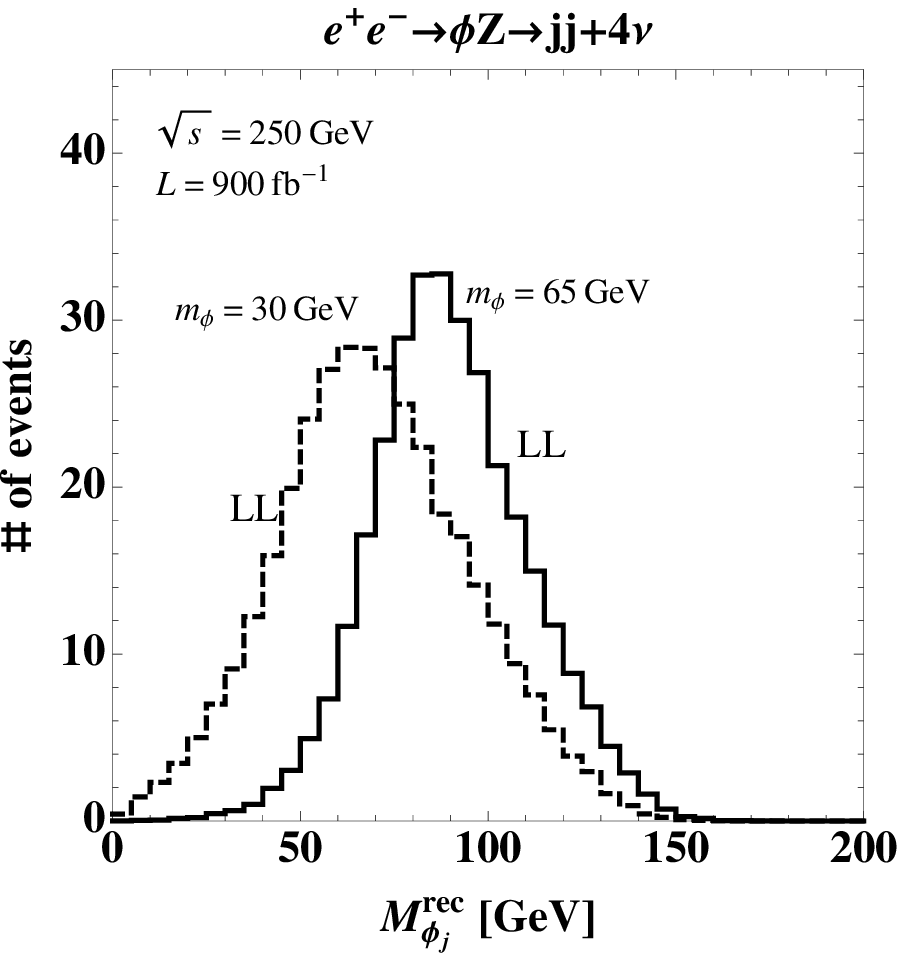} \quad
\includegraphics[width=50mm]{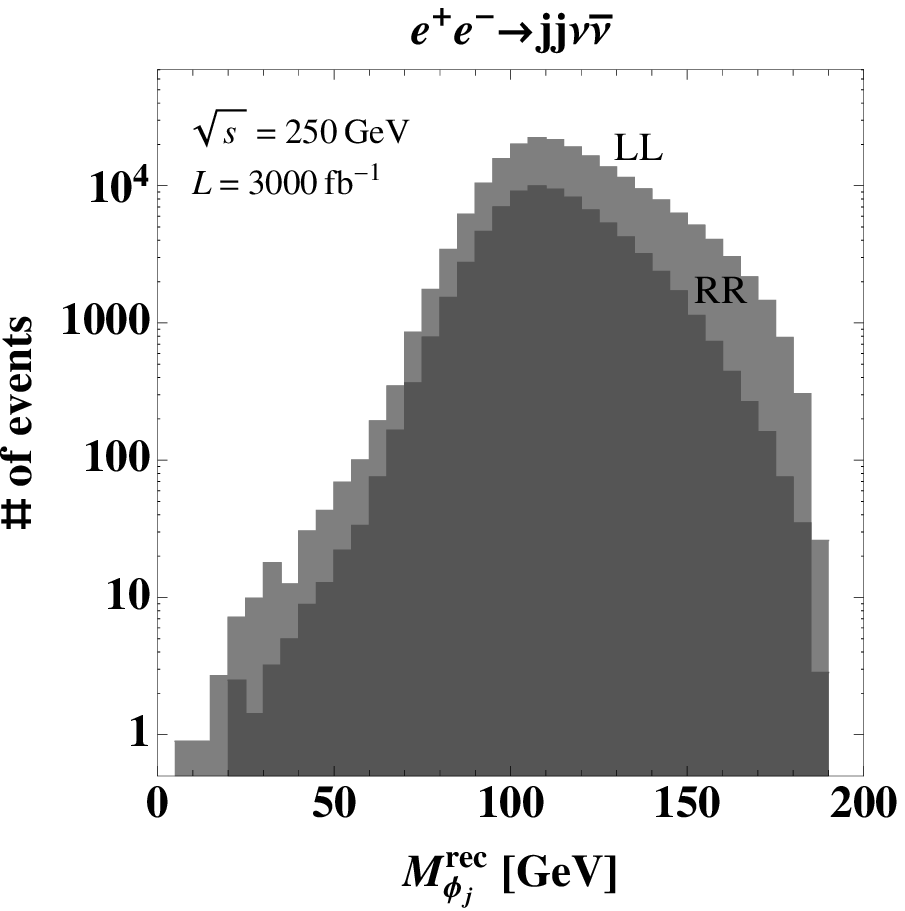} \quad
\includegraphics[width=50mm]{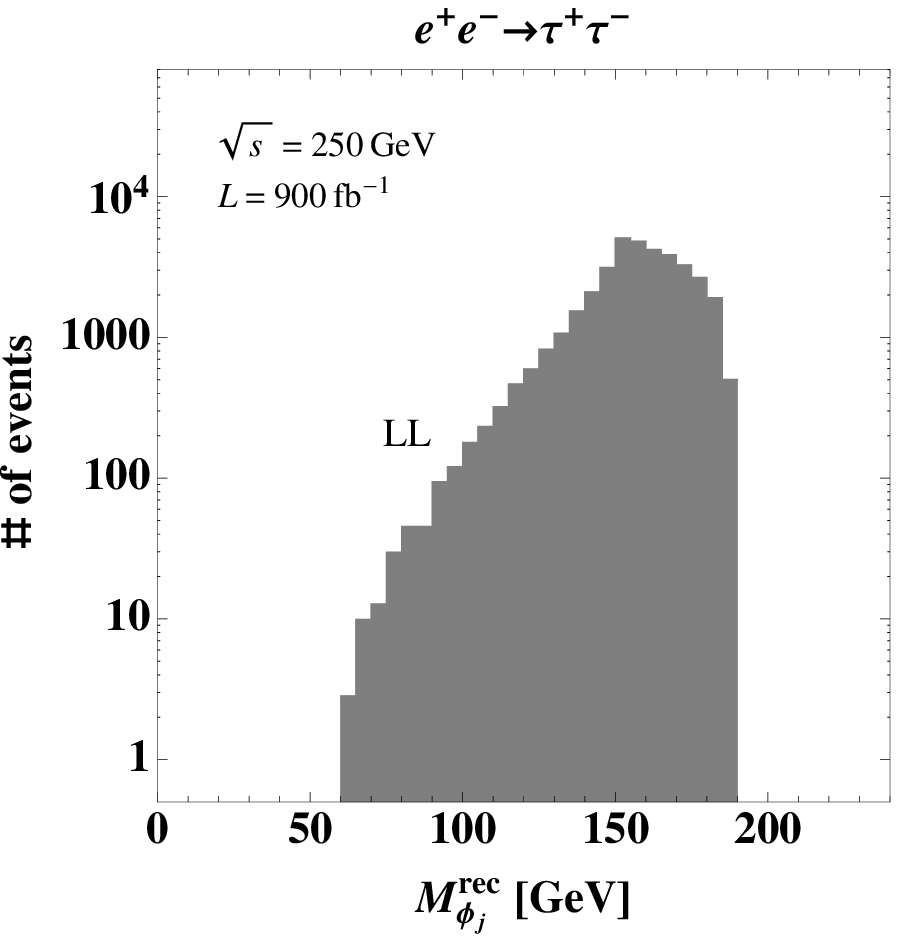} 
\caption{Distribution of reconstructed $\varphi$ mass after imposing basic and $M_{jj}$ cuts where left-, middle- and right-panels correspond to signal, $jj \nu \bar \nu$ BG and $\tau^+ \tau^-$ BG events.  Here $\kappa_\alpha =1$ is applied.} 
  \label{fig:MphiDistjj}
\end{center}\end{figure}
As in the $\ell^+\ell^- + \slashed{E}$ case, we investigate kinematical distributions for the signal and BGs to find relevant kinematical cuts.
Plots in Fig.~\ref{fig:MjjDist} show distributions of invariant mass of two jets where left-, middle- and right-panels correspond to events from signal, $jj\nu \bar \nu$ BG and $\tau^+ \tau^-$ BG with only basic cuts in Eq.~(\ref{eq:BasicCutsJ}). 
To compare with "$\ell^+ \ell^- + \slashed{E}$" case we show distribution for both $LL$ and $RR$ polarization in $jj \nu \bar \nu$ BG, and we find the behaviors are not significantly different in these polarizations since the BG comes from $ZZ$ production; the distributions for the other plots have also similar behavior in $LL$ and $RR$ polarizations.
The distribution for signal shows $Z$ peak which is slightly broader than that in $\ell^+\ell^-$ case above and the position of peak is slightly smaller than $Z$ boson mass; this is due to the fact that jet energy resolution is worse than that of charged leptons. The $jj \nu \bar \nu$ BG case also shows distribution peaked around $Z$ boson mass. The distribution for $\tau^+ \tau^-$ BG shows broad bump peaked around 160 GeV. In reducing BG events, we thus impose $jj$ invariant mass cuts such that
\begin{equation}
m_Z - 20 \ {\rm GeV} < M_{jj} < m_Z + 5 \ {\rm GeV}.
\end{equation}
We also reconstruct the mass of $\phi$ as in the case of charged lepton final state with energy momentum conservation.
Similarly we obtain the reconstructed mass as 
\begin{equation}
M_{\phi_j}^{rec} = \sqrt{s + m_Z^2 -2 (E_{j_1} + E_{j_2}) \sqrt{s} }
\end{equation}
where $E_{j_i}$ is energy of a jet in final state $j_i$.
Plots in Fig.~\ref{fig:MphiDistjj} show the distribution of $M_{\phi_j}^{rec}$ for signal and BGs.
We see that the reconstructed mass of $\phi$ tends to larger than actual value of $m_\phi$ and peak for $jj \nu \bar \nu$ BG is also larger than $m_Z$.
This is due to energy loss of two jets due to initial/final state radiation which is stronger than the case of charged lepton final states. 
Then we impose kinematical cuts for $M_{\phi_j}^{rec}$ such that
\begin{equation}
m_\phi - 15(10) \ {\rm GeV} < M_{\phi_j}^{rec} < m_\phi + 25(50) \ {\rm GeV},
\end{equation}
for $m_\phi = 65(30)$ GeV.
Table~\ref{tab:cut_effect_jjR} summarizes the effect of kinematical cuts to signal and BGs for $RR$ polarization.
We find that $\tau^+ \tau^-$ BG is highly suppressed by $M_{jj}$ and $M_{\phi_j}^{rec}$ cuts, and main BG after cuts is $jj \nu \bar \nu$ one.

Finally we estimate the discovery significance using Eq.~(\ref{eq:sig}) which is shown in the last column of Table~\ref{tab:cut_effect_jjR} for $RR$ polarization.
In addition, for comparison, we show significances for $RR$, $LL$ and sum of $LL$ and $RR$ polarizations in Table~\ref{tab:cut_effect_jjLR} for $m_{\phi} = 65(30)$ GeV. 
Significance tends to higher than that of "$\ell^+ \ell^- + \slashed{E}$" case; this is due to the facts that higher number of signal events by $BR(Z \to jj) > BR(Z \to \ell^+ \ell^-)$ 
and $e^+ e^- \to W^+W^-$ process does not contribute to $jj \nu \bar \nu$ final state. We then obtain significance much larger than 5 for $m_\phi = 30$ GeV with $\kappa_\alpha =1$ 
corresponding to $\sin \alpha =0.05$; $S_{cl} \sim 5$ can be obtained with $\sin \alpha =0.04$. 
Note also that we have the largest significance when we sum up events from $LL$ and $RR$ polarizations simply due to increase of the number of signal events.

\begin{center} 
\begin{table}[t]
\begin{tabular}{|c | c| c| c |c |}\hline
& $\kappa_\alpha N_{S}^{\kappa_\alpha=1}$; $m_\phi = (65, 30)$ GeV & \quad $N_{BG}^{jj \nu \bar \nu}$  \quad & \quad $N_{BG}^{\tau\tau}$ \quad & \quad $\kappa_\alpha S_{cl}^{\kappa_\alpha=1}$ \quad \\ \hline
Only basic cuts & ($3.8 \times 10^2$, $1.2 \times 10^3$)  & $1.1 \times 10^5$ &  $6.1 \times 10^5$ & (0.45, 0.46)   \\ 
+ $M_{jj}$ cut & ($2.9 \times 10^2$, $9.3 \times 10^2$)  & $8.0 \times 10^4$  &  $3.0 \times 10^4$ & (0.88, 1.1)  \\ \hline \hline
+ $M_{\phi_j}^{rec}$ cut for $m_\phi=65$ GeV & ($1.3 \times 10^2$,$\cdots$) & $5.7 \times 10^3$ & $1.3 \times 10^2$ & (1.6, $\cdots$) \\ 
+ $M_{\phi_j}^{rec}$ cut for $m_\phi=30$ GeV & ($\cdots$, $1.5 \times 10^2$) & $3.3 \times 10^2$ & 6.4 & ($\cdots$, 8.3) \\ \hline
\end{tabular}
\caption{The number of events for signal ($N_S$), BG ($N_{BG}$) and significance ($S_{cl}$) for $RR$ polarization after each cut where the setting is the same as Table.~\ref{tab:cut_effectR}}
\label{tab:cut_effect_jjR}
\end{table}
\end{center}

\begin{center} 
\begin{table}
\begin{tabular}{|c | c| c| c |c |}\hline
& $\kappa_\alpha N_{S}^{\kappa_\alpha=1}$; $m_\phi = 65(30)$ GeV & \quad $N_{BG}^{jj \nu \bar \nu}$  \quad & \quad $N_{BG}^{\tau\tau}$ \quad & \quad $\kappa_\alpha S_{cl}^{\kappa_\alpha=1}$ \quad \\ \hline
$RR$ & $1.3 (1.5) \times 10^2$  & $5.6 (0.33) \times 10^3$ & $1.3 (0.064) \times 10^2$  &  $1.6 (8.3)$  \\ 
 $LL$ & $1.6(1.9) \times 10^2$ &  $1.3  (0.085) \times 10^4$ & $2.0 (0.13) \times 10^2 $  & $1.4 (6.5)$   \\ \hline \hline
$LL + RR$ & $2.9 (3.4) \times 10^2$ & $1.9 (0.12) \times 10^4$ & $3.3 (0.19) \times 10^2$ & $2.1 (9.7)$ \\  \hline
\end{tabular}
\caption{The number of events for signal ($N_S$), BG ($N_{BG}$) and significance ($S_{cl}$) for $RR$ and $LL$ polarizations with integrated luminosity of $900$ fb$^{-1}$ each and for sum of events from two polarizaitons, where we show cases for $m_{\phi} = 65(30)$ GeV with all kinematical cuts imposed.}
\label{tab:cut_effect_jjLR}
\end{table}
\end{center}

Before closing this section, let us discuss the potential of the other lepton colliders and possibility of testing scalar mixing in future Higgs measurement. 
In addition to the ILC, the CEPC~\cite{CEPC} and FCC-ee~\cite{Gomez-Ceballos:2013zzn, FCC-ee} can investigate our scenario; the CEPC at $\sqrt{s} = 240$ GeV can provide data with integrated luminosity of 5 ab$^{-1}$ while at the FCC-ee integrated luminosity can be 10(5) ab$^{-1}$ for $\sqrt{s} = 160(\sim 250)$ GeV and that of 1.5 ab$^{-1}$ is possible for $\sqrt{s}= 350$ GeV. 
Then these experiments also have the potential to find the signature of our model which would give similar significance as our analysis since the energy and integrated luminosity are not significantly different from the case of the ILC. Thus combining the analysis of these experiments we can further improve the test of our model.
Moreover the lepton colliders can significantly improve measurements of the SM Higgs coupling which can constrain the scalar mixing. The couplings of $h ZZ$ interaction can be measured with the most strong sensitivity of $\sim 0.1 \%$ error and the other coupling can be also measured with few $\%$ error in each future lepton colliders~\cite{CEPC, Gomez-Ceballos:2013zzn,Fujii:2017vwa}. 
In our case, the SM Higgs coupling is given by $\cos \alpha \times C_{{h VV/ h\bar f f}}^{SM}$ where $C_{hVV/ h\bar f f}^{SM}$ is the SM Higgs coupling with vector bosons/fermions. 
Thus divination from the SM is given by $1 - \cos \alpha \simeq 0.0013 \times (\sin \alpha/0.05)^2$ which would be tested by $hZZ$ coupling measurement. 
The more stringent constraint can be obtained from future measurement of invisible decay branching ratio of the SM Higgs. For example, the ILC at $\sqrt{s}=250$ GeV with integrated luminosity of 2 ab$^{-1}$ can explore the branching ratio up to $0.32 \%$~\cite{Fujii:2017vwa}. Therefore, comparing with Fig.~\ref{fig:higgs-invisible}, wide parameter region can be explored which will be good complimentary test of our model.

\section{Summary and discussion}
We have studied a model with $U(1)_{L_\mu - L_\tau}$ gauge symmetry which is spontaneously broken by a VEV of SM singlet scalar field with non-zero $L_\mu - L_\tau$ charge. 
In this model $Z'$ boson and new CP-even scalar boson $\phi$ are obtained after spontaneous symmetry breaking.
Then we have focused on parameter region which can explain muon $g-2$ by one-loop contribution where $Z'$ boson propagates inside a loop, taking into account current experimental constraints. 
In the parameter region $Z'$ mass range is 5 MeV $\lesssim m_{Z'} \lesssim$ 210 MeV, and mass of $\phi$ is typically $\mathcal{O}(100)$ GeV. 
We have also found that $\phi$ dominantly decays into $Z'Z'$ mode and $Z'$ decays into $e^+e^-$ or $\bar \nu_\ell \nu_\ell$ modes depending on the ratio between $U(1)_{L_\mu - L_\tau}$ gauge coupling constant and kinetic mixing parameter.

Then we have investigated signatures of $\phi$ production processes in collider experiments.
Firstly gluon fusion production of $\phi$ at the LHC has been discussed considering mixing between the SM Higgs boson and $\phi$; 
the cross section is thus proportional to $\sin^2 \alpha$ with mixing angle $\alpha$.
In principle we can obtain sizable number of events from $pp \to \phi \to Z' Z'$ followed by decay of $Z' \to e^+ e^-$ even if Higgs-$\phi$ mixing is as small as $\sin \alpha \lesssim 0.1$.
However $e^+e^-$ pair from light $Z'$ decay is highly collimated and it is very challenging to analyze the signal events at the LHC requiring improved technology.

Secondly we have investigated $\phi$ production at $e^+ e^-$ collider such as the ILC.
In $e^+ e^-$ collider, $\phi$ can be produced via $e^+e^- \to Z \phi$, $W$ boson fusion and $Z$ boson fusion processes through the mixing with the SM Higgs boson.
Among them $Z \phi$ mode can give the largest cross section if kinematically allowed and we have focused on the process.
One advantage of $e^+e^-$ collider compared with hadron colliders is that we can use energy momentum conservation and $\phi$ mass can be reconstructed even if final state includes missing energy.  
In addition, we can use polarized electron/positron beam at the ILC experiment.
We have then considered  the process $e^+e^- \to Z \phi$ where $\phi$ decays into missing energy as $\phi \to Z' Z' \to 4 \nu$ since $BR(Z' \to \nu \bar \nu) \gg BR(Z' \to e^+ e^-)$ in the parameter region to give sizable muon $g-2$. For $Z$ boson decay, we have discussed two cases (1) $Z \to \ell^+ \ell^- (\ell = e, \mu)$ and (2) $Z \to j j$ giving "$\ell^+ \ell^+ +\slashed{E}$" and "$jj + \slashed{E}$" signal events respectively.
Numerical simulation study has been carried out for these cases generating signal events and the SM background events.
In our analysis, we have applied two polarization case in which $(e^-, e^+)$ beams are polarized as $(- 80 \%, + 30 \%)$ and $(+ 80 \%, - 30 \%)$ denoted by $LL$ and $RR$ polarizations respectively.
We have investigated relevant kinematical cuts to reduce the backgrounds showing corresponding distributions.  
Finally we have estimated discovery significance for our signal taking into account the effects of kinematical cuts.
The significance of $2.2(2.5)$ has been obtained for "$\ell^+ \ell^+ +\slashed{E}$" signal when we take $\sin \alpha=0.05$, $m_\phi =65(30)$ GeV and integrated luminosity of 900 fb$^{-1}$ for $RR$ polarization. Remarkably, we have the largest significance from $RR$ polarization which is even larger than sum of $LL$ and $RR$ events since BG from $e^+ e^- \to W^+ W^- \to \ell^+ \ell^- \nu \bar \nu$ process is suppressed in $RR$ polarization.
Furthermore the significance of $2.4(9.7)$ has been obtained for "$j j +\slashed{E}$" signal when we take $\sin \alpha=0.05$ and $m_\phi =65(30)$ GeV, which is larger than the case with charged lepton final state. 
In this case, we have find the largest significance can be obtained by simply summing up events from events $LL$ and $RR$ polarization.
In addition, we can obtain larger significance for larger $\sin \alpha$ although muon $g-2$ tends to become smaller.
Therefore we can search for the signal of $\phi$ at $e^+e^-$ collider with sufficient integrated luminosity, and
combining together with results from future muon $g-2$ measurements our $U(1)_{L_{\mu}- L_{\tau}}$ model will be further tested. 
Note also that the significance would be improved by more sophisticated cuts and further analysis will be given elsewhere. 
For the last comment, we discuss displaced vertex of $Z'$ decay into $e^+ e^-$. From Eq.~(20), 
the order of the lifetime can be estimated as $\tau_{Z'}  \simeq 24 \pi/(g'^2 m_{Z'}) \sim 4 \times 10^{-14}$ sec.,
where we assumed $g' = 10^{-4}$ and $m_{Z'} = 100$ MeV. The decay length is $c \tau_{Z'} \sim 1$ cm 
which is comparable with the radius of an innermost vertex tracker at the ILC. Therefore displaced vertices of $Z'$ 
decaying into $e^+ e^-$ might be measured if enough number of $Z'$ is produced. 

\section*{Acknowledgments}
This work is supported by JSPS KAKENHI Grants 
No.~15K17654 and 18K03651 (T.S.).
The authors would like to thank Hideki Okawa and Shin-ichi kawada for the private discussion.

\end{document}